\documentclass[useAMS,usegraphicx,usenatbib]{mn2e}

\usepackage{times}

\newcommand{\Msun}{\ensuremath{\,{\rm M}_\odot}}                  
\newcommand{\Rsun}{\ensuremath{\,{\rm R}_\odot}}                  
\newcommand{\Teff}{\ensuremath{T_{\rm eff}}}                      
\newcommand{\logg}{\ensuremath{\log g}}                           
\newcommand{\Mjup}{\ensuremath{\,{\rm M}_{\rm Jup}}}              
\newcommand{\Rjup}{\ensuremath{\,{\rm R}_{\rm Jup}}}              
\newcommand{\Teq}{\ensuremath{T_{\rm eq}^{\,\prime}}}             
\newcommand{\safronov}{\ensuremath{\Theta}}                       
\newcommand{\ms}{\,m\,s$^{-1}$}                                   
\newcommand{\mss}{\,m\,s$^{-2}$}                                  
\newcommand{\Porb}{\ensuremath{P_{\rm orb}}}                      
\newcommand{\as}{\ensuremath{^{\prime\prime}}}                    
\newcommand{\am}{\ensuremath{^\prime}}                            
\newcommand{\FeH}{\ensuremath{\left[{\rm Fe}/{\rm H}\right]}}     
\newcommand{\pjup}{\ensuremath{\,\rho_{\rm Jup}}}                 
\newcommand{\psun}{\ensuremath{\,\rho_\odot}}                     
\newcommand{\chir}{\ensuremath{\chi_\nu^{\,2}}}                   
\newcommand{\mc}[1]{\multicolumn{2}{c}{#1}}
\newcommand{\mcc}[1]{\multicolumn{3}{c}{#1}}
\newcommand{\er}[3]{\ensuremath{#1^{+#2}_{-#3}}}
\newcommand{\erc}[3]{\mc{\ensuremath{#1^{+#2}_{-#3}}}}

\newcommand{\reff}[1]{{#1}}

\title[The WASP-57 planetary system]
      {Larger and faster: revised properties and a shorter orbital period for the WASP-57 planetary system from a pro-am collaboration\thanks{Based on data collected by MiNDSTEp with the Danish 1.54\,m telescope, and data collected with GROND on the MPG 2.2\,m telescope, both located at ESO La Silla.}}

\author[Southworth et al.]
       {John Southworth$^{1}$,
        L.\ Mancini\,$^{2}$,
        J.\ Tregloan-Reed\,$^{3}$,
        S.\ Calchi Novati\,$^{4,5,6}$,
        S.\ Ciceri\,$^{2}$,
        \newauthor
        G.\ D'Ago\,$^{6,5,7}$,
        L.\ Delrez\,$^{8}$,
        M.\ Dominik\,$^{9}$,
        D.\ F.\ Evans\,$^{1}$,
        M.\ Gillon\,$^{8}$,
        E.\ Jehin\,$^{8}$,
        \newauthor
        U.\ G.\ J{\o}rgensen\,$^{10}$,
        T.\ Haugb{\o}lle\,$^{10}$,
        M.\ Lendl\,$^{8,11}$,
        C.\ Arena\,$^{12,13}$,
        L.\ Barbieri\,$^{12,14}$,
        \newauthor
        M.\ Barbieri\,$^{15}$,
        G.\ Corfini\,$^{12}$,
        C.\ Lopresti\,$^{12,16}$,
        A.\ Marchini\,$^{17,12}$,
        G.\ Marino\,$^{12,18}$,
        \newauthor
        K.\ A.\ Alsubai\,$^{19}$,
        V.\ Bozza\,$^{5,7}$,
        D.\ M.\ Bramich\,$^{19}$,
        R.\ Figuera Jaimes$^{9,20}$,
        T.\ C.\ Hinse\,$^{21}$,
        \newauthor
        Th.\ Henning\,$^{2}$,
        M.\ Hundertmark\,$^{10}$,
        D.\ Juncher\,$^{10}$,
        H.\ Korhonen\,$^{22,10}$,
        A.\ Popovas\,$^{10}$,
        \newauthor
        M.\ Rabus\,$^{23,2}$,
        S.\ Rahvar\,$^{24}$,
        R.\ W.\ Schmidt\,$^{25}$,
        J.\ Skottfelt\,$^{26,10}$,
        C.\ Snodgrass\,$^{27}$,
        \newauthor
        D.\ Starkey\,$^{9}$,
        J.\ Surdej\,$^{8}$,
        O.\ Wertz\,$^{8}$
        \\
        $^{1}$\,Astrophysics Group, Keele University, Staffordshire, ST5 5BG, UK \\
        $^{2}$\,Max Planck Institute for Astronomy, K\"onigstuhl 17, 69117 Heidelberg, Germany \\
        $^{3}$\,NASA Ames Research Center, Moffett Field, CA 94035, USA \\
        $^{4}$\,NASA Exoplanet Science Institute, MS 100-22, California Institute of Technology, Pasadena, CA 91125, US \\
        $^{5}$\,Dipartimento di Fisica ``E.R. Caianiello'', Universit\`a di Salerno, Via Giovanni Paolo II 132, 84084, Fisciano (SA), Italy \\
        $^{6}$\,Istituto Internazionale per gli Alti Studi Scientifici (IIASS), 84019 Vietri Sul Mare (SA), Italy \\
        $^{7}$\,Istituto Nazionale di Fisica Nucleare, Sezione di Napoli, 80126 Napoli, Italy \\
        $^{8}$\,Institut d'Astrophysique et de G\'eophysique, Universit\'e de Li\`ege, 4000 Li\`ege, Belgium \\
        $^{9}$\,SUPA, University of St Andrews, School of Physics \& Astronomy, North Haugh, St Andrews, KY16 9SS, UK \\
        $^{10}$\,Niels Bohr Institute \& Centre for Star and Planet Formation, University of Copenhagen, {\O}ster Voldgade 5, 1350 Copenhagen, Denmark \\
        $^{11}$\,Observatoire de Gen\`eve, Universit\'e de Gen\`eve, Chemin des maillettes 51, 1290 Sauverny, Switzerland \\
        $^{12}$\,Sezione Pianeti Extrasolari -- Unione Astrofili Italiani (UAI) \\
        $^{13}$\,Gruppo Astrofili Catanesi -- Catania, Italy \\
        $^{14}$\,AAB Associazione Astrofili Bolognesi, Bologna, Italy \\
        $^{15}$\,Department of Physics, University of Atacama, Copayapu 485, Copiapo, Chile \\
        $^{16}$\,Istituto Spezzino Ricerche Astronomiche -- I.R.A.S., La Spezia, Italy \\
        $^{17}$\,DSFTA -- Astronomical Observatory, University of Siena, Via Roma 56, 53100 Siena, Italy \\
        $^{18}$\,Gruppo Astrofili Catanesi (GAC) -- Catania, Italy \\
        $^{19}$\,Qatar Environment and Energy Research Institute, Qatar Foundation, Tornado Tower, Floor 19, P.O.\ Box 5825, Doha, Qatar \\
        $^{20}$\,European Southern Observatory, Karl-Schwarzschild-Stra{\ss}e 2, 85748 Garching bei M\"unchen, Germany \\
        $^{21}$\,Korea Astronomy and Space Science Institute, Daejeon 305-348, Republic of Korea \\
        $^{22}$\,Finnish Centre for Astronomy with ESO (FINCA), University of Turku, V{\"a}is{\"a}l{\"a}ntie 20, FI-21500 Piikki{\"o}, Finland \\
        $^{23}$\,Instituto de Astrof{\i}sica, Facultad de F{\i}sica, Pontificia Universidad Cat\'olica de Chile, Av.\ Vicu\~na Mackenna 4860, 7820436 Macul, Santiago, Chile \\
        $^{24}$\,Department of Physics, Sharif University of Technology, P.\,O.\,Box 11155-9161 Tehran, Iran \\
        $^{25}$\,Astronomisches Rechen-Institut, Zentrum f\"ur Astronomie, Universit\"at Heidelberg, M\"onchhofstra{\ss}e 12-14, 69120 Heidelberg, Germany \\
        $^{26}$\,Centre of Electronic Imaging, Department of Physical Sciences, The Open University, Milton Keynes, MK7 6AA, UK \\
        $^{27}$\,Planetary and Space Sciences, Department of Physical Sciences, The Open University, Milton Keynes, MK7 6AA, UK
        }

\begin{document} \maketitle 


\begin{abstract}
Transits in the WASP-57 planetary system have been found to occur half an hour earlier than expected. We present ten transit light curves from amateur telescopes, on which this discovery was based, thirteen transit light curves from professional facilities which confirm and refine this finding, and high-resolution imaging which show no evidence for nearby companions. We use these data to determine a new and precise orbital ephemeris, and measure the physical properties of the system. Our revised orbital period is 4.5\,s shorter than found from the discovery data alone, which explains the early occurrence of the transits. We also find both the star and planet to be larger and less massive than previously thought. The measured mass and radius of the planet are now consistent with theoretical models of gas giants containing no heavy-element core, as expected for the sub-solar metallicity of the host star. Two transits were observed simultaneously in four passbands. We use the resulting light curves to measure the planet's radius as a function of wavelength, finding that our data are sufficient in principle but not in practise to constrain its atmospheric properties. We conclude with a discussion of the current and future status of transmission photometry studies for probing the atmospheres of gas-giant transiting planets.
\end{abstract}

\clearpage



\section{Introduction}                                                                                                              \label{sec:intro}

Although the first transiting extrasolar planet (TEP) was only discovered in late 1999 \citep{Henry+00apj,Charbonneau+00apj}, and the second as recently as 2003 \citep{Konacki+03nat}, the number currently known has already exceeded 1200\footnote{See TEPCat (Transiting Extrasolar Planet Catalogue; \citealt{Me11mn}) at: \ {\tt http://www.astro.keele.ac.uk/jkt/tepcat/}}. The great majority of those are small objects observed using the NASA {\it Kepler} satellite: validation of the planetary nature of these bodies has been greatly helped by their occurrence in systems of multiple planets \citep[see][]{Rowe+14apj} but detailed studies are difficult due to their small size and long orbital periods (\Porb s).

A significant number (231 as of 2015/07/21) of the known TEPs are hot Jupiters, adopting a definition of mass $M_{\rm b} > 0.3$\Mjup\ and $\Porb < 10$\,d. These are much better suited to characterisation with existing facilities, as their relatively large masses and radii, short orbital periods, and bright host stars make photometric and spectroscopic observations easier and more productive. Perhaps the single most important observable property of a planet is its orbital period: the period distributions of exoplanets provide an insight into the mechanisms governing their formation and evolution \citep[e.g.][]{Mordasini+09aa,Mordasini+09aa2,Benitez++11aa,Hellier+12mn}, and a precise value is mandatory for performing follow-up observations.

\begin{table*} \centering
\caption{\label{tab:amateur} Instrumental setup for the amateur observations. $N_{\rm obs}$ is the number of observations.}
\begin{tabular}{llllll} \hline
Date       & Observer     & Telescope                               & CCD             & Filter      & $N_{\rm obs}$ \\
\hline
2014/05/24 & C.\ Lopresti & 180\,mm Maksutov-Newton                 & SBIG ST10XME    & red         & 39            \\
2014/06/10 & G.\ Corfini  & 200\,mm aperture, 800\,mm focal length  & SBIG STT-1603   & clear       & 87            \\
2014/06/10 & C.\ Lopresti & 300\,mm aperture, 1500\,mm focal length & SBIG ST10XME    & red         & 42            \\
2014/06/10 & A.\ Marchini & 300\,mm Zen Maksutov-Cassegrain         & STL-6303        & Cousins $R$ & 58            \\
2014/06/27 & L.\ Barbieri & 300\,mm aperture, 3000\,mm focal length & SBIG ST9        & clear       & 40            \\
2014/06/27 & G.\ Corfini  & 200\,mm aperture, 800\,mm focal length  & SBIG STT-1603   & clear       & 34            \\
2014/06/27 & C.\ Lopresti & 180\,mm Maksutov-Newton                 & SBIG ST10XME    & red         & 16            \\
2014/06/27 & C.\ Lopresti & 300\,mm aperture, 1500\,mm focal length & SBIG ST10XME    & red         & 58            \\
2014/06/27 & G.\ Marino   & 250\,mm aperture, 1200\,mm focal length & SBIG ST7-XME    & clear       & 64            \\
2015/05/28 & C.\ Lopresti & 180\,mm Maksutov-Newton                 & SBIG ST10XME    & red         & 30            \\
\hline \end{tabular} \end{table*}

\begin{table*} \centering
\caption{\label{tab:obslog} Log of the observations obtained from professional telescopes. $N_{\rm obs}$ is the
number of observations, $T_{\rm exp}$ is the exposure time, $T_{\rm dead}$ is the dead time between exposures,
`Moon illum.' is the fractional illumination of the Moon at the midpoint of the transit, and $N_{\rm poly}$
is the order of the polynomial fitted to the out-of-transit data. The aperture radii are target aperture,
inner sky and outer sky, respectively. The `bb' filter is a blue-blocking filter.}
\setlength{\tabcolsep}{4pt}
\begin{tabular}{lcccccccccc@{~}c@{~}ccc} \hline
Instrument & Date of   & Start time & End time  &$N_{\rm obs}$ & $T_{\rm exp}$ & $T_{\rm dead}$ & Filter & Airmass & Moon & \mcc{Aperture radii}   & $N_{\rm poly}$ & Scatter \\
           & first obs &    (UT)    &   (UT)    &              & (s)           & (s)            &        &         &illum.& \mcc{(pixels)} &                & (mmag)  \\
\hline
TRAPPIST    & 2012/03/15 & 04:34 & 08:20 & 367 &    20   & 10 & bb  & 1.83 $\to$ 1.12 $\to$ 1.13 & 0.463 & 11.3 & 22.7 & 36.3 & 1 & 2.25 \\ 
TRAPPIST    & 2012/04/01 & 04:40 & 09:36 & 701 &    15   &  7 & bb  & 1.36 $\to$ 1.12 $\to$ 1.45 & 0.649 & 13.5 & 19.3 & 30.9 & 1 & 3.49 \\ 
Euler       & 2012/04/01 & 05:20 & 09:35 & 212 & 50--180 & 16 & $r$ & 1.23 $\to$ 1.12 $\to$ 1.44 & 0.649 & 24   &      &      & 0 & 0.98 \\ 
BUSCA       & 2012/05/10 & 23:16 & 02:49 &  82 &   120   & 35 & $u$ & 1.31 $\to$ 1.29 $\to$ 1.86 & 0.665 & 15   & 25   & 45   & 1 & 3.46 \\ 
BUSCA       & 2012/05/10 & 23:16 & 03:37 & 100 &   120   & 35 & $g$ & 1.31 $\to$ 1.29 $\to$ 2.46 & 0.665 & 17   & 27   & 50   & 1 & 1.26 \\ 
BUSCA       & 2012/05/10 & 23:16 & 03:34 &  99 &   120   & 35 & $r$ & 1.31 $\to$ 1.29 $\to$ 2.44 & 0.665 & 18   & 28   & 50   & 1 & 0.82 \\ 
BUSCA       & 2012/05/10 & 23:16 & 03:37 &  98 &   120   & 35 & $z$ & 1.31 $\to$ 1.29 $\to$ 2.46 & 0.665 & 18   & 28   & 50   & 1 & 1.58 \\ 
DFOSC       & 2014/05/18 & 01:58 & 06:51 & 162 &   100   &  9 & $R$ & 1.32 $\to$ 1.12 $\to$ 1.70 & 0.761 & 12   & 20   & 40   & 1 & 0.74 \\ 
DFOSC       & 2014/06/24 & 23:33 & 04:15 & 155 &   100   &  8 & $R$ & 1.27 $\to$ 1.12 $\to$ 1.53 & 0.050 & 16   & 26   & 45   & 1 & 0.71 \\ 
GROND       & 2014/06/24 & 23:38 & 02:34 &  53 & 90--110 & 54 & $g$ & 1.25 $\to$ 1.12 $\to$ 1.17 & 0.050 & 40   & 60   & 90   & 1 & 0.89 \\ 
GROND       & 2014/06/24 & 23:38 & 02:34 &  53 & 90--110 & 54 & $r$ & 1.25 $\to$ 1.12 $\to$ 1.17 & 0.050 & 40   & 60   & 90   & 1 & 0.58 \\ 
GROND       & 2014/06/24 & 23:38 & 02:34 &  53 & 90--110 & 54 & $i$ & 1.25 $\to$ 1.12 $\to$ 1.17 & 0.050 & 32   & 55   & 75   & 1 & 0.58 \\ 
GROND       & 2014/06/24 & 23:38 & 02:34 &  53 & 90--110 & 54 & $z$ & 1.25 $\to$ 1.12 $\to$ 1.17 & 0.050 & 30   & 50   & 80   & 1 & 1.02 \\ 
\hline \end{tabular} \end{table*}


In this work we study the WASP-57 system, whose planetary nature was discovered by the SuperWASP consortium \citep[][hereafter F13]{Faedi+13aa}. WASP-57 contains a star slightly cooler and less massive than the Sun ($\Teff = 5600 \pm 100$\,K, $M_{\rm A} = 0.89 \pm 0.07$\Msun) orbited by a planet which is the same size but less massive than Jupiter ($M_{\rm b} = 0.64 \pm 0.06$\Mjup, $R_{\rm b} = 1.05 \pm 0.05$\Rjup). The moderately different properties found by F13 placed the planet at the lower edge of the distribution of gas giant TEPs in the mass--radius diagram, making it a good candidate for hosting a heavy-element core despite the subsolar metallicity of the host star ($\FeH = -0.25 \pm 0.10$). The analysis by F13 was based on SuperWASP photometry \citep{Pollacco+06pasp}, radial velocities from CORALIE spectra, plus two complete and one partial transit light curves from the Euler and TRAPPIST telescopes at ESO La Silla, Chile. No further work on this system has been published.

Early in the 2014 observing season a group of amateur astronomers noticed that the transits of WASP-57 were occurring half an hour earlier than expected, which is a significant fraction of the 2.3\,hr total transit duration. This was immediately confirmed using a transit of WASP-57 which had been serendipitously observed on 2014/05/18 using the Danish 1.5\,m telescope, at La Silla. A second transit observation was scheduled on 2014/06/24 with the Danish telescope, the ESO 2.2\,m telescope and GROND imager, and the immediately following transit with the amateur observers in Europe, allowing its early arrival to be reconfirmed. We also possess high-precision light curves of WASP-57 obtained in the 2012 season, before the imprecision of the original orbital ephemeris became apparent. In this work we present all the data we have obtained for WASP-57, produce a revised ephemeris which can be used for follow-up observations in future, measure the physical properties of the system to high precision, and search for variations of the measured planetary radius with wavelength


\section{Observations and data reduction}                                                                                             \label{sec:obs}

\begin{figure*} \includegraphics[width=\textwidth,angle=0]{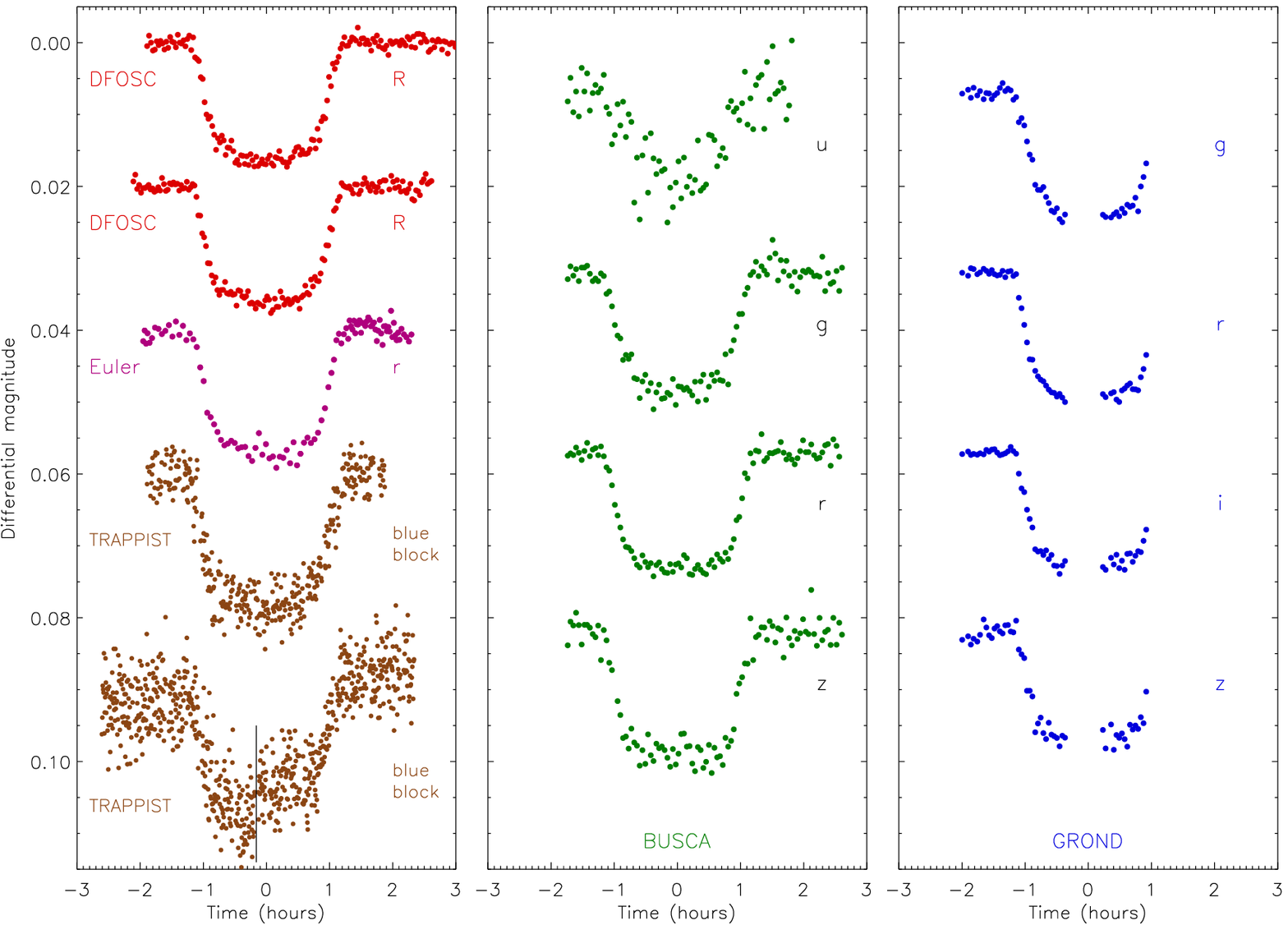}
\caption{\label{fig:lc:all} All light curves from professional facilities
presented in this work, grouped and colour-coded according to the telescope
used. The instrument and filter are labelled individually for each light curve.
The second light curve from TRAPPIST has a discontinuity shortly before the
midpoint of the transit due to a meridian-flip. \reff{This is indicated
using a vertical black line.}} \end{figure*}

A total of ten transit light curves were obtained by LB, GC, CL, AM and GM using telescopes of apertures between 180\,mm and 300\,mm, sited in Italy. Further details of the observational setup and numbers of datapoints are given in Table\,\ref{tab:amateur}.

Two complete transits of WASP-57 were observed using the 1.54\,m Danish Telescope and DFOSC instrument at ESO La Silla, Chile \citep[see][]{Dominik+10an}, on the dates 2014/05/18 and 2014/06/24. DFOSC has a plate scale of 0.39\as\,pixel$^{-1}$ and a 2048$^2$ pixel CCD, giving a field of view of 13.7\am$\times$13.7\am. We windowed the CCD down to 1100$\times$900 and 1045$\times$920 pixels to shorten the dead time between exposures, resulting in images containing WASP-57 and six decent comparison stars. Both transits were obtained through a Bessell $R$ filter. The instrument was defocussed in order to improve the efficiency of the observations, and to combat time-correlated noise \citep[see][]{Me+09mn}. The telescope was autoguided to limit pointing drifts to less than five pixels over each observing sequence. An observing log is given in Table\,\ref{tab:obslog} and the light curves are plotted individually in Fig.\,\ref{fig:lc:all}.

The transit on 2014/06/24 observed with DFOSC was also monitored using GROND \citep{Greiner+08pasp} mounted on the MPG 2.2\,m telescope at La Silla, Chile. GROND was used to obtain light curves simultaneously in four passbands, which approximate the SDSS $g$, $r$, $i$ and $z$ bands. The small field of view of this instrument (5.4$^{\prime}$$\times$5.4$^{\prime}$ at a plate scale of 0.158$^{\prime\prime}$\,pixel$^{-1}$) meant that fewer comparison stars were available. The instrument was defocussed and the telescope was autoguided throughout the observing sequence. Poor weather conditions (high wind) forced closure of the telescope before the transit finished, so the light curves have only partial coverage of the transit (see Table\,\ref{tab:obslog} and Fig.\,\ref{fig:lc:all}).

We now turn to observations obtained prior to the conception of the current work. We observed WASP-57 on the night of 2012/05/10 using the BUSCA instrument on the 2.2\,m telescope at Calar Alto Astronomical Observatory. BUSCA is capable of observing simultaneously in four passbands, for which we chose Gunn $u$, $g$, $r$ and $z$. The motivation for these choices, and a detailed discussion on the use of BUSCA for planetary transit observations, can be found in \citet{Me+12mn2}. All four CCDs on BUSCA have a plate scale of 0.176\as\ pixel$^{-1}$, but were operated with 2$\times$2 binning. Whilst the full field of view of 12$\times$12 arcmin was accessible in the $u$ band, the available field in the $g$, $r$ and $z$ bands was vignetted into a circle of diameter of approximately 6\am. The instrument was defocussed and the telescope was autoguided throughout the observations (Table\,\ref{tab:obslog}). The $u$-band data are of insufficient quality for full modelling but can be used to obtain a time of minimum and to check for a possible variation of measured planetary radius with wavelength.

One transit of WASP-57 was observed on 2012/04/01 with EulerCam, using the same methods as for the EulerCam transit in F13. EulerCam is a CCD imager mounted on the 1.2\,m Euler-Swiss telescope, La Silla, with a field of view of 14.7\am$\times$14.7\am\ at 0.23\as\,pixel$^{-1}$. We obtained 212 images through a Gunn $r$ filter, without applying a defocus to the instrument. Further details on EulerCam and the data reduction proceduce can be found in \citet{Lendl+12aa}.

Two transits of WASP-57 were observed on 2012/03/15 and 2012/04/01 using the 0.6\,m TRAPPIST robotic telescope located at La Silla \citep{Gillon+11conf,Jehin+11msngr}. The 2k$\times$2k CCD was thermoelectrically cooled and yielded a field of view of 22\am$\times$22\am\ at 0.65\as\,pixel$^{-1}$. Images were obtained with a slight defocus and through a blue-blocking filter\footnote{{\tt http://www.astrodon.com/products/filters/exoplanet/}} that has a transmittance greater than 90\% from 500\,nm to beyond 1000\,nm.

\subsection{Data reduction}

The data from the amateur telescopes were all reduced using MaxIm DL\footnote{{\tt http://www.cyanogen.com/maxim\_main.php}}. In each case the science images were calibrated using dark and flat-field frames.

The data from DFOSC, GROND and BUSCA were reduced using aperture photometry as implemented in the {\sc defot} code \citep{Me+09mn,Me+14mn}, which relies on the {\sc idl\footnote{The acronym {\sc idl} stands for Interactive Data Language and is a trademark of ITT Visual Information Solutions. For further details see: {\tt http://www.exelisvis.com/ProductsServices/IDL.aspx}.}/astrolib\footnote{The {\sc astrolib} subroutine library is distributed by NASA. For further details see: {\tt http://idlastro.gsfc.nasa.gov/}.}} implementation of {\sc daophot} \citep{Stetson87pasp}. Master bias and flat-field images were constructed but generally found to have an insignificant effect on the quality of the photometry. Image motion was tracked by cross-correlating individual images with a reference image.

We obtained photometry on the instrumental system using software apertures of a range of sizes, and retained those which gave light curves with the smallest scatter (Table\,\ref{tab:obslog}). We found that the choice of aperture size affects the scatter but not the shape of the transit in the final light curve. The instrumental magnitudes were then transformed to differential-magnitude light curves, normalised to zero magnitude outside transit using first-order polynomials (Table\,\ref{tab:obslog}) fitted to the out-of-transit data. The differential magnitudes are relative to a weighted ensemble of typically five (DFOSC) or two to four (GROND) comparison stars. The comparison star weights and polynomial coefficients were simultaneously optimised to minimise the scatter in the out-of-transit data.

Finally, the timestamps for the datapoints were converted to the BJD(TDB) timescale \citep{Eastman++10pasp}. We performed manual time checks for several images obtained with DFOSC and verified that the FITS file timestamps are on the UTC system to within a few seconds. In recent work on the WASP-103 system we found that the timestamps from DFOSC and GROND agree to within a few seconds, supporting the reliability of both \citep{Me+15mn}. The reduced data are given in Table\,\ref{tab:lcdata} and will be lodged with the CDS\footnote{{\tt http://vizier.u-strasbg.fr/}}.

The data from EulerCam were reduced using aperture photometry following the methods given by \citet{Lendl+12aa}. Differential aperture photometry was also used on the TRAPPIST data, using carefully selected extraction apertures and reference stars. For more details on the TRAPPIST data reduction procedures, see e.g.\ \citet{Gillon+13aa}. The transit on 2012/04/01 was obtained in two sequences separated by a meridian flip, and the two sets of data were reduced independently.

\begin{table} \centering \caption{\label{tab:lcdata} Sample of the
data presented in this work (the first and last datapoints of each
light curve). The full dataset will be made available at the CDS.}
\setlength{\tabcolsep}{4pt}
\begin{tabular}{l c r@{.}l r@{.}l r@{.}l} \hline
Instrument & Filter & \mc{BJD(TDB)} & \mc{Diff.\ mag.} & \mc{Uncertainty} \\
\hline
TRAPPIST & bb  & 2456001&690740 &  0&0014100 & 0&0033979 \\
TRAPPIST & bb  & 2456001&847510 &  0&0018100 & 0&0019776 \\[2pt]
TRAPPIST & bb  & 2456018&694550 &  0&0062000 & 0&0041227 \\
TRAPPIST & bb  & 2456018&900580 & -0&0042000 & 0&0041094 \\[2pt]
Euler    & $r$ & 2456018&722460 &  0&0015300 & 0&0014096 \\
Euler    & $r$ & 2456018&899430 &  0&0006000 & 0&0010572 \\[2pt]
BUSCA    & $u$ & 2456058&476778 &  0&0011721 & 0&0027719 \\
BUSCA    & $u$ & 2456058&624676 & -0&0073014 & 0&0050536 \\[2pt]
BUSCA    & $g$ & 2456058&476778 &  0&0008973 & 0&0010736 \\
BUSCA    & $g$ & 2456058&657776 & -0&0006933 & 0&0015483 \\[2pt]
BUSCA    & $r$ & 2456058&476778 &  0&0004552 & 0&0007219 \\
BUSCA    & $r$ & 2456058&655976 &  0&0005960 & 0&0009239 \\[2pt]
BUSCA    & $z$ & 2456058&476778 &  0&0018345 & 0&0015082 \\
BUSCA    & $z$ & 2456058&657776 &  0&0003519 & 0&0017916 \\[2pt]
DFOSC    & $R$ & 2456796&588810 &  0&0004779 & 0&0006752 \\
DFOSC    & $R$ & 2456796&792254 &  0&0006067 & 0&0008685 \\[2pt]
DFOSC    & $R$ & 2456833&486396 & -0&0006900 & 0&0007120 \\
DFOSC    & $R$ & 2456833&683083 & -0&0007523 & 0&0007212 \\[2pt]
GROND    & $g$ & 2456833&490392 &  0&0000938 & 0&0008559 \\
GROND    & $g$ & 2456833&611846 &  0&0097990 & 0&0009372 \\[2pt]
GROND    & $r$ & 2456833&490392 &  0&0000128 & 0&0005593 \\
GROND    & $r$ & 2456833&611846 &  0&0114375 & 0&0006182 \\[2pt]
GROND    & $i$ & 2456833&490392 &  0&0002314 & 0&0005610 \\
GROND    & $i$ & 2456833&611846 &  0&0107411 & 0&0009272 \\[2pt]
GROND    & $z$ & 2456833&490392 &  0&0010646 & 0&0010125 \\
GROND    & $z$ & 2456833&611846 &  0&0082877 & 0&0010501 \\
\hline \end{tabular} \end{table}

\subsection{High-resolution imaging}                                                                                               \label{sec:obs:li}

\begin{figure} \centering
\includegraphics[width=0.9\columnwidth,angle=0]{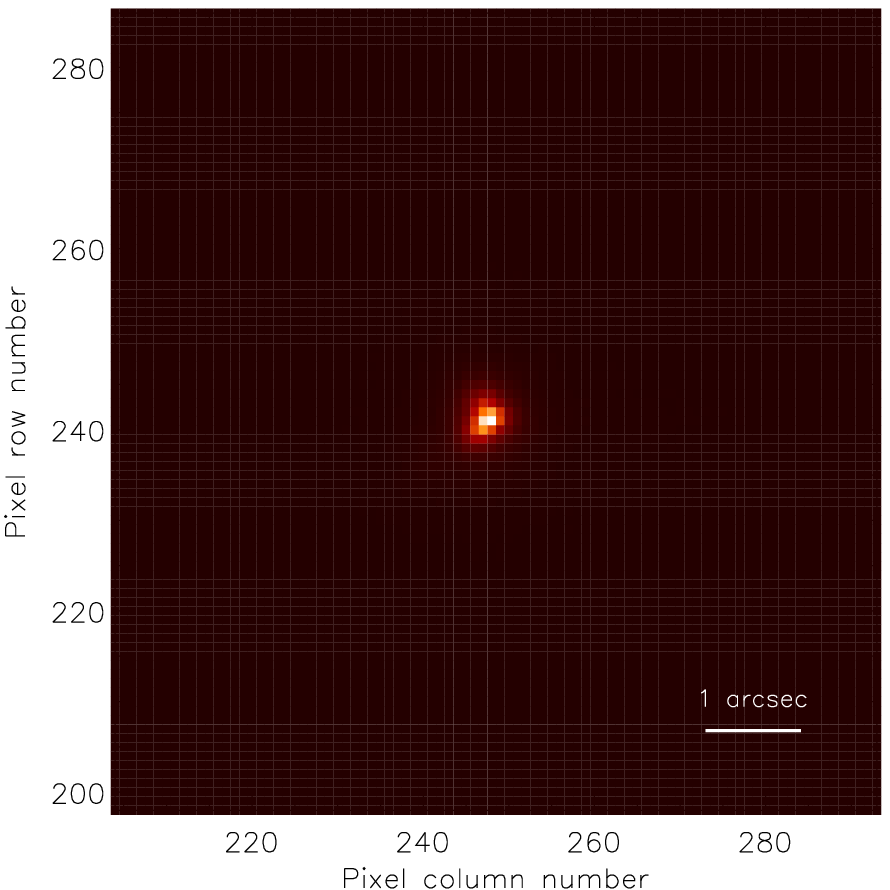}
\includegraphics[width=0.9\columnwidth,angle=0]{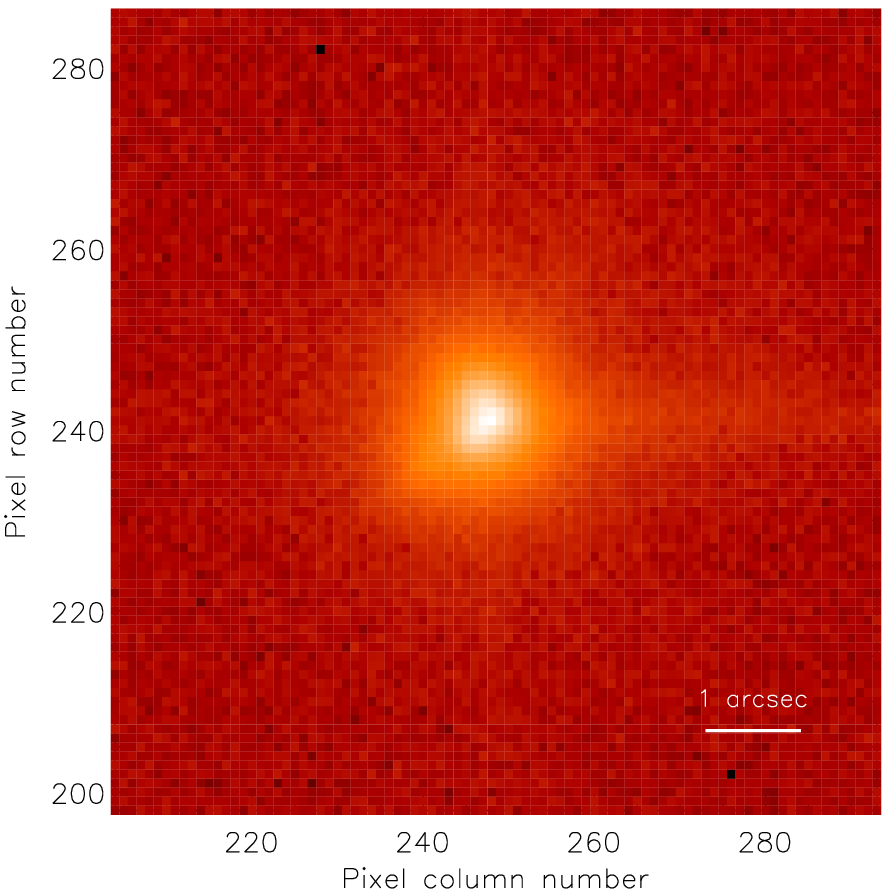}
\caption{\label{fig:li} High-resolution Lucky Image of the field around
WASP-57. The upper panel has a linear flux scale for context and the lower
panel has a logarithmic flux scale to enhance the visibility of any faint
stars. Each image covers $8\as \times 8\as$ centred on WASP-57. A bar of
length $1\as$ is superimposed in the bottom-right of each image. The image
is a sum of the best 2\% of the original images.} \end{figure}

\begin{figure}
\includegraphics[width=\columnwidth,angle=0]{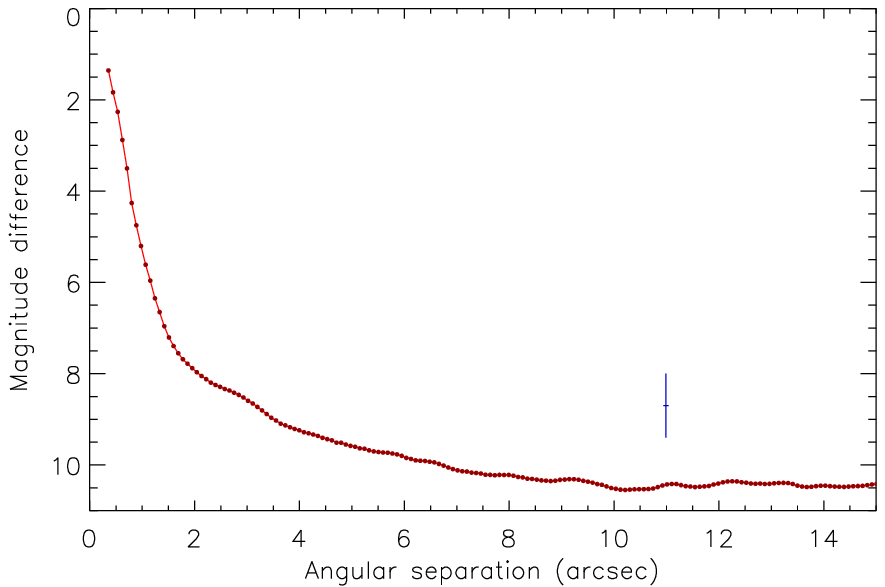}
\caption{\label{fig:contrast} \reff{Constrast curve giving the limiting
magnitude of the LI observation as a function of angular distance
from WASP-57 (dark red circles connected by a red line). The closest
detected star is shown as a blue datapoint.}} \end{figure}

We obtained several images of WASP-57 with DFOSC in sharp focus, allowing us to check for the presence of faint nearby stars whose light might act to decrease the observed transit depth \citep{Daemgen+09aa}. \reff{The closest stars we found on any image are much fainter than WASP-57, and are over 45\as\ distant, so are too far away to affect our photometry.}

We also obtained a high-resolution image of WASP-57 using the Lucky Imager (LI) mounted on the Danish telescope \citep[see][]{Skottfelt+13aa,Skottfelt+15aa}. The LI uses an Andor 512$\times$512 pixel electron-multiplying CCD, with a pixel scale of 0.09\as\,pixel$^{-1}$ giving a field of view of $45\as\times45\as$. The data were reduced using a dedicated pipeline and the 2\% of images with the smallest point spread function (PSF) were shifted and added to yield a combined image whose PSF is smaller than the seeing limit. A long-pass dichroic was used, resulting in a response function which approximates that of SDSS $i$$+$$z$. An overall exposure time of 600\,s corresponds to an effective exposure time of 12\,s for the best 2\% of the images (Fig.\,\ref{fig:li}). The FWHM of the PSF is 4.0$\times$4.2 pixels ($0.36\as \times 0.38\as$).

\reff{Two faint stars were detected on the LI image, at angular distances of $10.99 \pm 0.05$\as\ and $21.56 \pm 0.07$\as\ from WASP-57, and fainter by $8.7 \pm 0.7$\,mag and $8.2 \pm 0.4$\,mag. Neither of these stars is sufficiently bright and close to WASP-57 to affect the analysis presented in the current work. We assessed the limiting magnitude of the LI image by placing a box with sides equal to the FWHM of the star on each pixel on the image. The standard deviation of the counts within each box was calculated, and a 3$\sigma$ detection threshold was generated. The mean detection threshold at a given radius from the target star was then converted to a relative magnitude. Further details of the detection and reduction methods are given in Evans et al.\ (2015, in preparation). The contrast curve is shown in Fig.\,\ref{fig:contrast}.}


\section{Transit timing analysis}                                                                                                    \label{sec:porb}

\begin{figure*}
\includegraphics[width=\textwidth,angle=0]{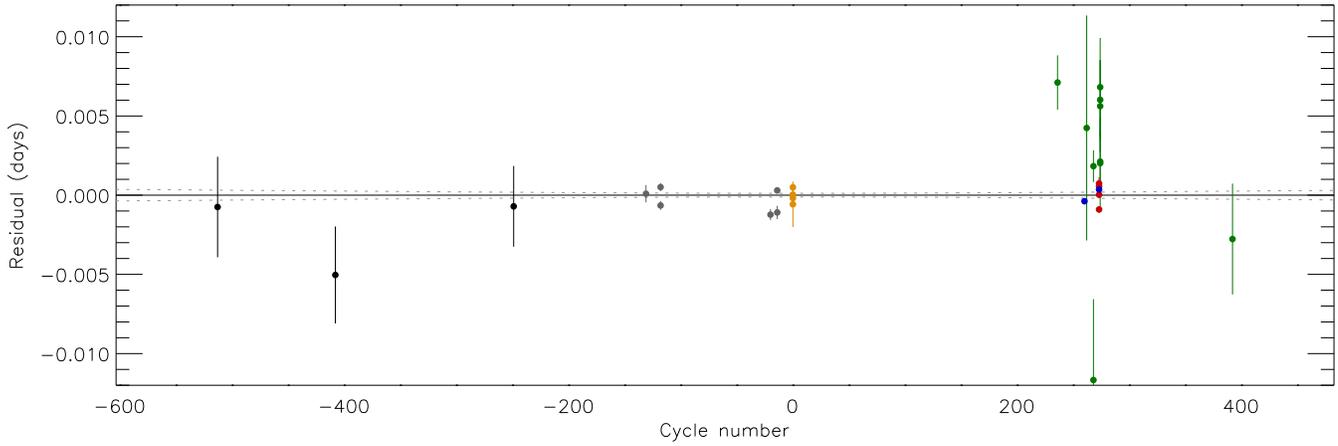}
\caption{\label{fig:minima} Plot of the residuals of the timings of mid-transit for
WASP-57 versus a linear ephemeris \reff{(see Table\,\ref{tab:minima})}. The points are colour-coded according to their
source: black for the WASP data, green for the amateur timings in the current work,
blue for DFOSC, red for GROND, off-yellow for BUSCA, and grey for the TRAPPIST and
Euler telescopes. The dotted lines show the 1$\sigma$ uncertainty in the ephemeris
as a function of cycle number.} \end{figure*}

\begin{table} \begin{center}
\caption{\label{tab:minima} Times of minimum light and their residuals versus the ephemeris derived in this work.
All but one of the timings were derived in the current work, from the source data given in the final column.}
\setlength{\tabcolsep}{3pt}
\begin{tabular}{l l r r l} \hline
Time of min. & Error & Cycle & Residual & Source \\
(BJD/TDB)  & (d) & no. & (d) &       \\
\hline
2454602.18313 & 0.00317 &  -513.0 & -0.00075 & This work (WASP 2008)  \\   
2454900.26529 & 0.00306 &  -408.0 & -0.00504 & This work (WASP 2009)  \\   
2455351.65767 & 0.00255 &  -249.0 & -0.00071 & This work (WASP 2010)  \\   
2455686.65086 & 0.00055 &  -131.0 &  0.00009 & This work (TRAPPIST)   \\   
2455723.55722 & 0.00025 &  -118.0 &  0.00051 & This work (Euler)      \\   
2455723.55606 & 0.00028 &  -118.0 & -0.00065 & This work (TRAPPIST)   \\   
2456001.76950 & 0.00035 &   -20.0 & -0.00123 & This work (TRAPPIST)   \\   
2456018.80454 & 0.00017 &   -14.0 &  0.00030 & This work (Euler)      \\   
2456018.80315 & 0.00042 &   -14.0 & -0.00109 & This work (TRAPPIST)   \\   
2456058.54852 & 0.00143 &     0.0 & -0.00058 & This work (BUSCA $u$)  \\   
2456058.54891 & 0.00039 &     0.0 & -0.00019 & This work (BUSCA $g$)  \\   
2456058.54959 & 0.00023 &     0.0 &  0.00049 & This work (BUSCA $r$)  \\   
2456058.54913 & 0.00042 &     0.0 &  0.00003 & This work (BUSCA $z$)  \\   
2456728.54099 & 0.00171 &   236.0 &  0.00711 & Dittler (TRESCA)       \\   
2456796.66754 & 0.00019 &   260.0 & -0.00038 & This work (DFOSC)      \\   
2456802.35000 & 0.00710 &   262.0 &  0.00424 & This work (Lopresti)   \\   
2456819.38110 & 0.00100 &   268.0 &  0.00183 & This work (Corfini)    \\   
2456819.36760 & 0.00510 &   268.0 & -0.01167 & This work (Marchini)   \\   
2456833.57422 & 0.00017 &   273.0 &  0.00036 & This work (DFOSC)      \\   
2456833.57442 & 0.00034 &   273.0 &  0.00056 & This work (GROND $g$)  \\   
2456833.57296 & 0.00021 &   273.0 & -0.00090 & This work (GROND $r$)  \\   
2456833.57389 & 0.00022 &   273.0 &  0.00003 & This work (GROND $i$)  \\   
2456833.57460 & 0.00039 &   273.0 &  0.00074 & This work (GROND $z$)  \\   
2456836.41480 & 0.00290 &   274.0 &  0.00202 & This work (LBarbieri)  \\   
2456836.41880 & 0.00250 &   274.0 &  0.00602 & This work (Corfini)    \\   
2456836.41490 & 0.00190 &   274.0 &  0.00212 & This work (Lopresti)   \\   
2456836.41960 & 0.00310 &   274.0 &  0.00682 & This work (Lopresti)   \\   
2456836.41840 & 0.00290 &   274.0 &  0.00562 & This work (Marino)     \\   
2457171.40240 & 0.00350 &   392.0 & -0.00277 & This work (Lopresti)   \\   
\hline \end{tabular} \end{center} \end{table}

The issue which brought WASP-57 to our attention was the offset between the predicted and actual times of transit. We have therefore obtained as many measured times of mid-transit as possible. We first modelled the two DFOSC transits individually using the {\sc jktebop} code (see below), as these are the two light curves which have full coverage of a transit with a low scatter in the data. We scaled the errorbars for each light curve to yield a reduced $\chi^2$ of $\chi^2_\nu = 1.0$ versus the fitted model. This step is necessary because the uncertainties from the {\sc aper} algorithm tend to be moderately too small.

We then modelled the light curves from the amateur telescopes with {\sc jktebop} but fitting for only the time of mid-transit and the out-of-transit brightness of the system. The other photometric parameters were fixed to the best-fitting values from the two DFOSC light curves. The uncertainties in the transit times were multiplied by $\sqrt{\chi^2_\nu}$ to account for the underestimated observational errors in most of the datasets. We performed the same process on the GROND data, except this time we fitted the out-of-transit brightness as a linear function of time rather than just a constant offset from zero differential magnitude.

We then turned to published data. The discovery paper of WASP-57 (F13) contains two light curves observed with TRAPPIST and one with the Euler telescope. We fitted these as above, with the photometric parameters fitted for the two light curves with complete transit coverage and fixed for the TRAPPIST light curve which only contains the second half of a transit. We also included one transit time obtained by U.\ Dittler and lodged on the Exoplanet Transit Database\footnote{The Exoplanet Transit Database (ETD) can be found at: {\tt http://var2.astro.cz/ETD/credit.php}; see also TRESCA at: {\tt http://var2.astro.cz/EN/tresca/index.php}.} \citep{Poddany++10newa}.

The SuperWASP data which triggered the discovery of the planetary nature comprise approximately 30\,000 datapoints obtained during the 2008, 2009 and 2010 observing seasons. These data were obtained and separated into individual seasons, then fitted with {\sc jktebop} in the same way as for the data obtained using amateur telescopes. The resulting season-averaged times of minimum are consistent with the linear ephemeris found below, but are of low precision. WASP-57\,A, at $V=13.04$, is comparatively faint for the SuperWASP telescopes so suffers from a large scatter in its light curve. We included these times of minimum light in the following analysis, but note that they do not have a significant effect on the results.

All times of mid-transit were then fitted with a straight line versus cycle number to determine a new linear orbital ephemeris. Table\,\ref{tab:minima} gives all transit times plus their residual versus the fitted ephemeris. We chose the reference epoch to be that for our BUSCA observations, in order to limit the covariance between the reference time of minimum and the orbital period. The resulting ephemeris is
$$ T_0 = {\rm BJD(TDB)} \,\, 2\,456\,058.54910 (16) \, + \, 2.83891856 (81) \times E $$
where $E$ gives the cycle count versus the reference epoch and the bracketed quantities indicate the uncertainty in the final digit of the preceding number. This orbital period is 4.5\,s (24$\sigma$) smaller than the value of 2.838971\,(2)\,d found by F13, explaining why we found the transits of WASP-57 to occur earlier than predicted. \reff{There are several plausible reasons for such a discrepancy to occur, but we are not in a position to choose between them.}

\begin{table*} \centering \caption{\label{tab:lcfit} Parameters of the
fit to the light curves of WASP-57 from the {\sc jktebop} analysis (top).
The final parameters are given in bold and the parameters found by F13 are
given below this. Quantities without quoted uncertainties were not given by
F13 but have been calculated from other parameters which were.
}
\begin{tabular}{l r@{\,$\pm$\,}l r@{\,$\pm$\,}l r@{\,$\pm$\,}l r@{\,$\pm$\,}l r@{\,$\pm$\,}l}
\hline
Source        & \mc{$r_{\rm A}+r_{\rm b}$} & \mc{$k$} & \mc{$i$ ($^\circ$)} & \mc{$r_{\rm A}$} & \mc{$r_{\rm b}$} \\
\hline
Euler     (2011/06/10) & 0.1198 & 0.0084 & 0.1139 & 0.0027 & 87.04 & 1.10 & 0.1075 & 0.0073 & 0.01225 & 0.00109 \\
TRAPPIST  (2011/06/10) & 0.1065 & 0.0065 & 0.1087 & 0.0027 & 89.94 & 1.36 & 0.0961 & 0.0055 & 0.01044 & 0.00073 \\
Euler     (2012/04/01) & 0.1256 & 0.0066 & 0.1190 & 0.0029 & 86.56 & 0.73 & 0.1123 & 0.0057 & 0.01335 & 0.00099 \\
TRAPPIST  (2012/03/15) & 0.1436 & 0.0095 & 0.1263 & 0.0026 & 85.28 & 0.84 & 0.1275 & 0.0082 & 0.01610 & 0.00136 \\
TRAPPIST  (2012/04/01) & 0.1405 & 0.0153 & 0.1179 & 0.0042 & 85.16 & 1.20 & 0.1257 & 0.0140 & 0.01482 & 0.00194 \\
BUSCA $g$ (2012/05/10) & 0.1370 & 0.0131 & 0.1188 & 0.0056 & 85.53 & 1.12 & 0.1224 & 0.0113 & 0.01455 & 0.00170 \\
BUSCA $r$ (2012/05/10) & 0.1340 & 0.0057 & 0.1186 & 0.0015 & 85.70 & 0.52 & 0.1198 & 0.0050 & 0.01421 & 0.00074 \\
BUSCA $z$ (2012/05/10) & 0.1328 & 0.0098 & 0.1230 & 0.0018 & 85.69 & 0.84 & 0.1182 & 0.0086 & 0.01454 & 0.00116 \\
DFOSC     (2014/05/18) & 0.1309 & 0.0057 & 0.1173 & 0.0018 & 86.04 & 0.52 & 0.1171 & 0.0049 & 0.01374 & 0.00077 \\
DFOSC     (2014/06/24) & 0.1306 & 0.0066 & 0.1166 & 0.0017 & 86.05 & 0.63 & 0.1170 & 0.0057 & 0.01364 & 0.00086 \\
\hline
Final results&{\bf0.1278}&{\bf0.0033}&{\bf0.1182}&{\bf0.0013}&{\bf86.05}&{\bf0.28}&{\bf0.1143}&{\bf0.0029}&{\bf0.01331}&{\bf0.00051}\\
\hline
F13 & \mc{0.1122} & 0.1127 & 0.0006 & \erc{88.0}{0.1}{0.2} & \mc{0.1008} & \mc{0.01135} \\
\hline \end{tabular} \end{table*}

The $\chi^2_\nu$ of the fit is 1.99, and we interpret this as an indication that the uncertainty estimates for the timings are too small. Fig.\,\ref{fig:minima} shows the residuals of the times of mid-transit versus the ephemeris given above. There is no sign of long-term transit timing variations.



\section{Light curve analysis}                                                                                                         \label{sec:lc}

Eight of our light curves cover a full transit at high photometric precision. The Euler and one of the two TRAPPIST light curves from F13 also satisfy this criterion. Each of these ten datasets was modelled separately using the {\sc jktebop}\footnote{{\sc jktebop} is written in {\sc fortran77} and the source code is available at {\tt http://www.astro.keele.ac.uk/jkt/codes/jktebop.html}} code \citep{Me++04mn} and the {\it Homogeneous Studies} methodology \citep[][and references therein]{Me12mn}. We did not subject those light curves with only partial coverage of a transit to this analysis, because the parameters derived from partial light curves are highly uncertain -- so have little effect on the final results -- and are often unreliable \citep[e.g.][]{Gibson+09apj}.

The {\sc jktebop} model is based on the fractional radii of the star and the planet ($r_{\rm A}$ and $r_{\rm b}$), which are the ratios between the true radii and the semimajor axis ($r_{\rm A,b} = \frac{R_{\rm A,b}}{a}$). The parameters of the fit to each light curve were the sum and ratio of the fractional radii ($r_{\rm A} + r_{\rm b}$ and $k = \frac{r_{\rm b}}{r_{\rm A}}$), the orbital inclination ($i$), limb darkening coefficients, and the time of mid-transit. We assumed an orbital eccentricity of zero, based on the finding by F13 that the \citet{LucySweeney71aj} test yielded a 100\% probability that the orbit was circular. We fixed the orbital period to the value found in Section\,\ref{sec:porb}. We also fitted for the coefficients of a first-order polynomial relating differential magnitude and time \citep{Me+14mn}, in order to allow for any errors in flux normalisation which change with time or airmass. The TRAPPIST light curve obtained on the night of 2012/04/01 was split into two sequences by a meridian flip. This was accounted for by modelling both sequences together but specifying a separate polynomial (of order 1) for each sequence.

Limb darkening (LD) was incorporated using each of five laws \citep[see][]{Me08mn}, with the linear coefficients either fixed at theoretically predicted values\footnote{Theoretical LD coefficients were obtained by bilinear interpolation in \Teff\ and \logg\ using the {\sc jktld} code available from: {\tt http://www.astro.keele.ac.uk/jkt/codes/jktld.html}} or included as fitted parameters. We did not calculate fits for both LD coefficients in the four two-coefficient laws as they are very strongly correlated
\citep{Carter+08apj}.
The nonlinear coefficients were instead perturbed by $\pm$0.1 on a flat distribution during the error analysis simulations, to account for the uncertainty in theoretical LD coefficients.

\begin{figure*} \includegraphics[width=\textwidth,angle=0]{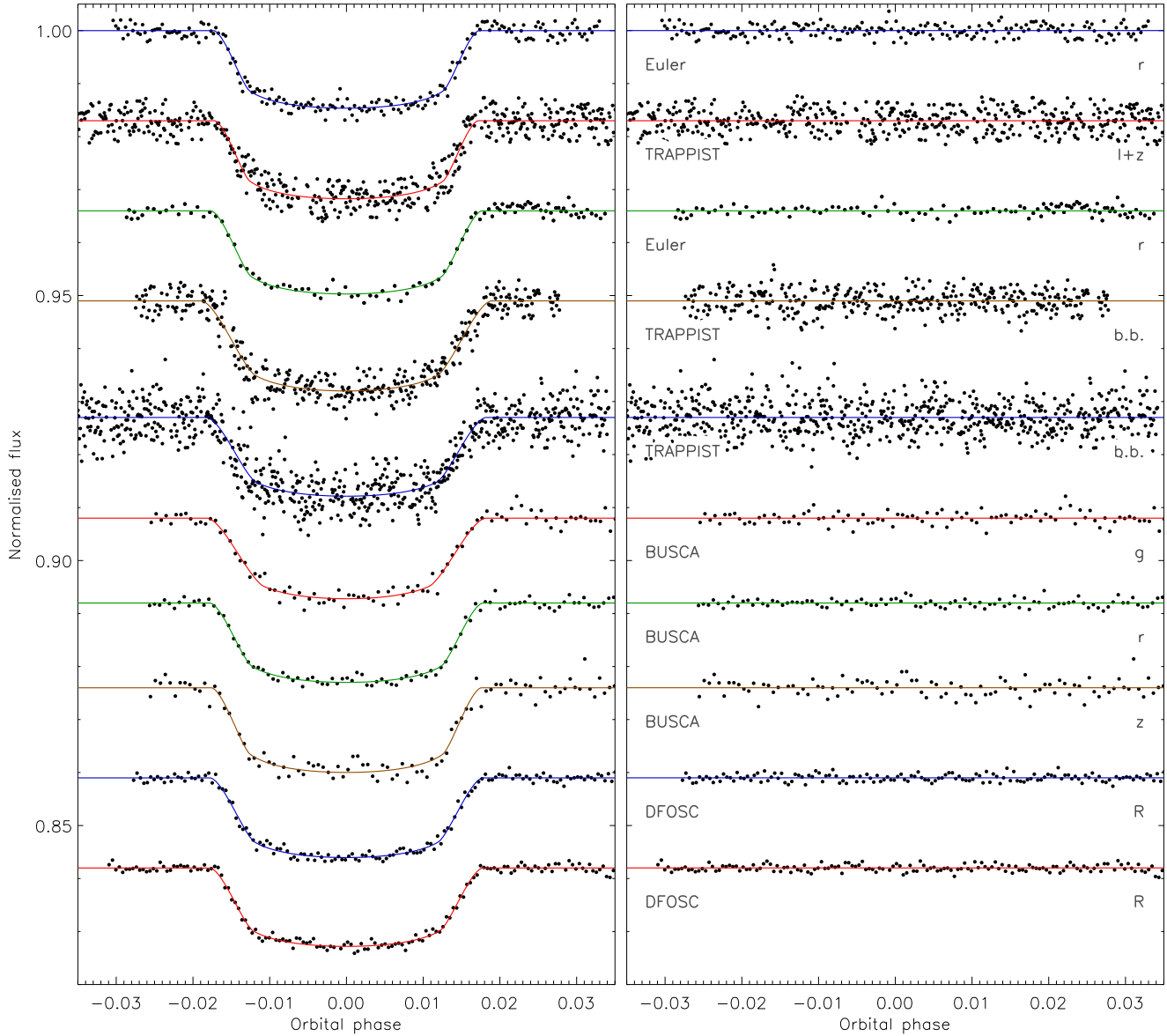}
\caption{\label{fig:lcfit} Phased light curves of WASP-57 compared to the
{\sc jktebop} best fits (left) and the residuals of the fits (right). Labels
give the source and passband for each dataset. The polynomial baseline functions
have been removed from the data before plotting. Only light curves with full
coverage of a transit were included in this analysis.} \end{figure*}

Error estimates for the fitted parameters were obtained in four steps. Steps 1 and 2 were residual-permutation and Monte Carlo simulations \citep{Me08mn}, and the larger of the two alternatives was retained for each fitted parameter. For step 3 we ran solutions using the five different LD laws, and increased the errorbar for each parameter to account for any disagreement between these five solutions. For step 4 we calculated the weighted mean of each photometric parameter using the values found separately from each light curve. This final step is a powerful external check on the reliability and mutual agreement between different datasets, as any discrepancies are obvious and quantifiable.

For all ten light curves we found that it was possible to fit for one of the two LD coefficients: reasonable values for the coefficients were obtained as well as a slightly smaller \chir\ compared to fits with both LD coefficients fixed. We therefore adopt these results, which are summarised in Table\,\ref{tab:lcfit}. Detailed tables of results for each light curve are available in the \reff{online-only} Appendix. The best fits are plotted in Fig.\,\ref{fig:lcfit}.



We find that the results from the different light curves are not in perfect agreement, with a \chir\ of 2.0 for $r_{\rm A}+r_{\rm b}$ and $r_{\rm A}$, 2.8 for $r_{\rm b}$ and 3.7 for $k$ versus the weighted mean of their values. This is due primarily to the TRAPPIST light curve from F13, which has a very small $r_{\rm A}$ and high $i$ compared to the other datasets (Table\,\ref{tab:lcfit}). A degeneracy between these parameters is common (e.g.\ \citealt{Carter+08apj} and \citealt{Me08mn}) and arises because these two values together specify the observed transit duration, a quantity which is well-determined by high-quality light curves. If we adopt instead the results from fitting this light curve with both LD coefficients fixed, the agreement becomes much better: $k$ has $\chir = 2.5$, the other four parameters in Table\,\ref{tab:lcfit} all have $\chir < 0.9$, and all photometric parameters change by less than their 1$\sigma$ errorbars. We have, however, chosen not to take this step for two reasons. Firstly, theoretical LD coefficients are not perfect -- if they were then different sources would give exactly the same values -- and none are available calculated specifically for the TRAPPIST $I$+$z$ filter. Secondly, taking an alternative approach for a discrepant dataset raises the possibility of causing an underestimate of the true uncertainties in the measured quantities. We have therefore retained the discrepant values when calculating the weighted means of the photometric parameters, and have inflated the errorbars on the weighted means by $\sqrt{\chir}$ in order to account for the discrepancy.

The photometric parameters found by F13 differ significantly from our results, by 4.7$\sigma$ for $r_{\rm A}$ and 3.8$\sigma$ for $r_{\rm b}$ (using our errorbars to calculate the $\sigma$ values as error estimates were not provided by F13 for these two quantities). This is due to the dependence of the F13 solution on only three transit light curves (two complete and one only partially covering a transit), all modelled simultaneously, of which one was the TRAPPIST dataset we find to be discrepant. The value of $k$ found by F13 ($0.1127 \pm 0.0006$) is 3.8$\sigma$ smaller than ours, and is evidence that the error estimates quoted by F13 are too small (see \citealt{Me12mn} and references therein for other examples). An alternative explanation is the presence of starspots, which is plausible for a star of this temperature. However, no traces of spot occultations are seen in our light curves and no rotational modulation is seen in the long-term WASP light curves.

\begin{table*} \centering \caption{\label{tab:model} Derived physical properties of WASP-57. The values found by F13 are given for comparison.}
\begin{tabular}{l l l r@{\,$\pm$\,}c@{\,$\pm$\,}l r@{\,$\pm$\,}l} \hline
Quantity                & Symbol           & Unit  & \mcc{This work}                 & \mc{F13}                       \\
\hline
Stellar mass            & $M_{\rm A}$      & \Msun & 0.886    & 0.061    & 0.028     & 0.954 & 0.028                  \\
Stellar radius          & $R_{\rm A}$      & \Rsun & 0.927    & 0.031    & 0.010     & \erc{0.836}{0.07}{0.16}        \\
Stellar surface gravity & $\log g_{\rm A}$ & \,cgs & 4.452    & 0.024    & 0.005     & \erc{4.574}{0.009}{0.012}      \\
Stellar density         & $\rho_{\rm A}$   & \psun & \mcc{$1.113 \pm 0.085$}         & \erc{1.638}{0.044}{0.063}      \\[2pt]
Planet mass             & $M_{\rm b}$      & \Mjup & 0.644    & 0.060    & 0.014     & \erc{0.672}{0.049}{0.046}      \\
Planet radius           & $R_{\rm b}$      & \Rjup & 1.050    & 0.052    & 0.011     & \erc{0.916}{0.017}{0.014}      \\
Planet surface gravity  & $g_{\rm b}$      & \mss  & \mcc{$14.5 \pm  1.5$}           & \erc{18.3}{2.9}{1.3}           \\
Planet density          & $\rho_{\rm b}$   & \pjup & 0.521    & 0.072    & 0.006     & \erc{0.873}{0.076}{0.071}      \\[2pt]
Equilibrium temperature & \Teq\            & K     & \mcc{$1338 \pm   29$}           & \erc{1251}{21}{22}             \\
Safronov number         & \safronov\       &       & 0.0522   & 0.0045   & 0.0006    & \mc{ }                         \\
Orbital semimajor axis  & $a$              & au    & 0.03769  & 0.00088  & 0.00040   & 0.0386 & 0.0004                \\
\hline \end{tabular} \end{table*}


\section{Physical properties}                                                                                                      \label{sec:absdim}

\begin{figure*} \includegraphics[width=\textwidth,angle=0]{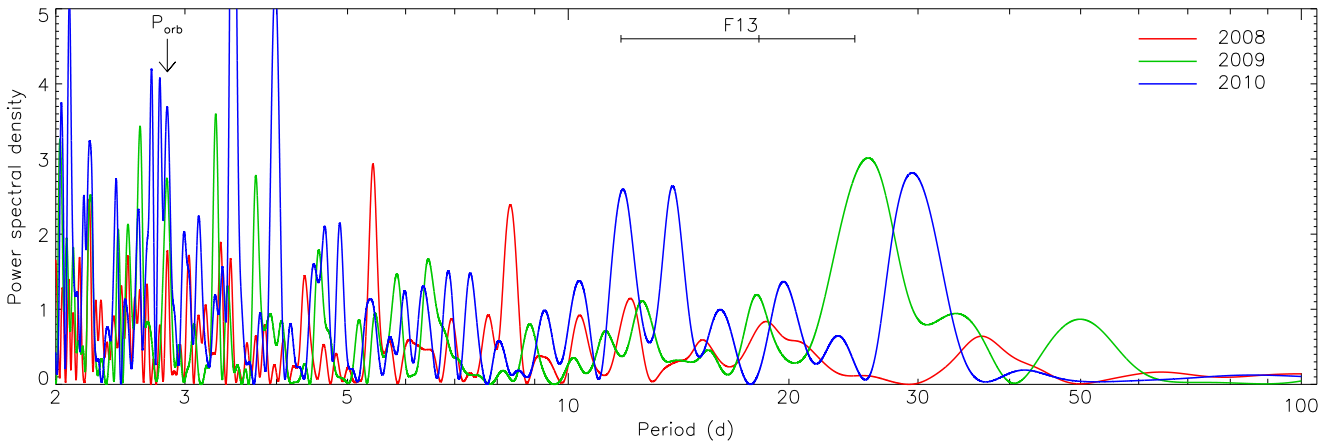}
\caption{\label{fig:psd} Scargle periodograms of the SuperWASP data from
the 2008, 2009 and 2010 seasons. The orbital period of the system is shown
with a downward-pointing arrow. The stellar rotational period inferred by
F13 from its projected rotational velocity is shown by the horizontal
errorbar.} \end{figure*}


We measured the physical properties of the WASP-57 system using the results from Section\,\ref{sec:lc}, five grids of predictions from theoretical models of stellar evolution \citep{Claret04aa,Demarque+04apjs,Pietrinferni+04apj,Vandenberg++06apjs,Dotter+08apjs}, and the host-star spectroscopic properties. Theoretical models provide an additional constraint on the stellar properties, needed because the system properties cannot be obtained from only measured quantities. The spectroscopic properties were obtained by F13 and comprise effective temperature ($\Teff = 5600 \pm 100$\,K), metallicity ($\FeH = -0.25 \pm 0.10$) and velocity amplitude ($K_{\rm A} = 100 \pm 7$\ms). We used the physical constants tabulated by \citet{Me11mn}.

We first estimated the velocity amplitude of the {\em planet}, $K_{\rm b}$, which was used along with the measured $r_{\rm A}$, $r_{\rm b}$, $i$ and $K_{\rm A}$ to determine the physical properties of the system \reff{\citep{Me09mn}}. The estimate of $K_{\rm b}$ was then iterated to find the best match between the measured $r_{\rm A}$ and the calculated $\frac{R_{\rm A}}{a}$, and the observed \Teff\ and that predicted by a theoretical model for the obtained stellar mass, radius and \FeH. This was done for a grid of ages from the zero-age main sequence to beyond the terminal-age main sequence for the star, in 0.01\,Gyr increments, and the overall best $K_{\rm b}$ was adopted. The statistical errors in the input quantities were propagated to the output quantities by a perturbation approach.

We ran the above analysis for each of the five sets of theoretical model predictions, yielding five different estimates of each output quantity. 
These were transformed into a single final result for each parameter by taking the unweighted mean of the five estimates and their statistical errors, plus an accompanying systematic error which gives the largest difference between the mean and individual values. The final results of this process are a set of physical properties for the WASP-57 system, each with a statistical error and a systematic error. The stellar density, planetary surface gravity and planetary equilibrium temperatures can be calculated without resorting to theoretical predictions \citep{SeagerMallen03apj,Me++07mn,Me10mn}, so do not have an associated systematic error.

\citet[][their tables 1 and 2]{Bodenheimer++03apj} provided predicted radii for planets of mass 0.69\Mjup\ and $\Teq = 1000$\,K and 1500\,K. Their radii are 1.01--1.10\Rjup\ with, and 1.03--1.13\Rjup\ without, a core or additional kinetic heating of the planetary interior. Both are in very good agreement with the radius of $1.05 \pm 0.05$\Rjup\ we find for WASP-57\,b.
\begin{table*} \centering
\caption{\label{tab:rb} \reff{Values of $r_{\rm b}$ and $R_{\rm b}$ for each of the light curves.
The errorbars exclude all common sources of uncertainty in $r_{\rm b}$ and $R_{\rm b}$ so should
only be used to compare different values of $r_{\rm b}(\lambda)$. The final column gives the size
of the errorbar on $R_{\rm b}$ in atmospheric scale heights.}}
\begin{tabular}{lllcr@{\,$\pm$\,}lr@{\,$\pm$\,}lr} \hline
Instrument & Passband & $\lambda_{\rm cen}$ & FWHM  & \mc{$r_{\rm b}$} & \mc{$R_{\rm b}$} & $\sigma$ ($H$) \\
           &          & (nm)                & (nm)  & \mc{ }           & \mc{(\Rjup)}     &     \\
\hline
Euler      & Gunn $r$     &  660.0  &  100.0  &  0.01362 & 0.00004  &   0.867 & 0.038   &  8.2  \\
BUSCA      & Gunn $u$     &  350.0  &   68.0  &  0.01099 & 0.00048  &   1.061 & 0.008   &  1.8  \\
BUSCA      & Gunn $g$     &  495.5  &   99.5  &  0.01345 & 0.00011  &   1.063 & 0.005   &  1.2  \\
BUSCA      & Gunn $r$     &  663.0  &  105.0  &  0.01348 & 0.00007  &   1.100 & 0.011   &  2.3  \\
BUSCA      & Gunn $z$     &  910.0  &   90.0  &  0.01395 & 0.00013  &   1.057 & 0.017   &  3.6  \\
DFOSC      & Bessell $R$  &  648.9  &  164.7  &  0.01343 & 0.00003  &   1.060 & 0.009   &  2.0  \\
GROND      & Gunn $g$     &  477.0  &  137.9  &  0.01340 & 0.00021  &   1.033 & 0.010   &  2.2  \\
GROND      & Gunn $r$     &  623.1  &  138.2  &  0.01344 & 0.00011  &   1.036 & 0.017   &  3.7  \\
GROND      & Gunn $i$     &  762.5  &  153.5  &  0.01310 & 0.00013  &   1.059 & 0.003   &  0.6  \\
GROND      & Gunn $z$     &  913.4  &  137.0  &  0.01314 & 0.00022  &   1.074 & 0.004   &  0.8  \\
\hline \end{tabular} \end{table*}

Table\,\ref{tab:model} contains our measurements of the physical properties of the WASP-57 system. Compared to F13, we find a less massive but larger star. As planetary properties are measured relative to those of their parent star, the planet is similarly affected. The measured planetary density is 3.5$\sigma$ lower, at $0.521 \pm 0.072$\pjup\ compared to the value of \er{0.873}{0.076}{0.071}\pjup\ found by F13. Our results are based on a much more extensive set of photometric data so are to be preferred to previous measurements, even though the errorbars have not changed by much. A significant advance in our understanding of the WASP-57 system could be achieved by obtaining \reff{further spectroscopy of the host star, from which more precise values for \Teff, \FeH\ and $K_{\rm A}$ could be measured}. This is of particular interest because of its metal-poor nature, whose effect on the incidence of different types of planets is currently under discussion \citep{Buchhave+14nat,WangFischer15aj}.

The age of the system is unconstrained in our analysis above, as often occurs when the host star is significantly less massive than 1\Msun. F13 inferred age estimates of $\goa$2\,Gyr for WASP-57\,A from its photospheric lithium abundance, and $\sim$$1.9^{+2.4}_{-1.2}$\,Gyr from gyrochronological arguments and its rotation period derived from its radius and projected rotational velocity. The \Teff\ of the star is within the regime where starspots are common so we have checked if it is possible to precisely determine its rotation period from spot-induced modulation. A Lomb-Scargle periodogram was calculated for each of the three seasons of SuperWASP data and can be seen in Fig.\,\ref{fig:psd}. There are no strong peaks in the period interval of interest (5--30\,d), and no moderately strong peaks present at the same period in all three seasons. We conclude that the rotational modulation of the star is below the level of detection with the current data.

\subsection{Comparison with theoretical models of giant planets}


\reff{F13 found that the measured mass and density of WASP-57\,b implied the presence of a heavy-element core of mass roughly 50\,M$_\oplus$, via a comparison to the theoretical predictions of \citet{Fortney++07apj}. This rather large core mass is surprising given the significantly sub-solar metal abundance of the host star ($\FeH = -0.25 \pm 0.10$). As we have found a significantly lower density for the planet (smaller by 40\% or 3.5$\sigma$) it is germane to reconsider this conclusion. We have therefore compared our new mass and radius measurements with predictions based on three batches of theoretical models.}

\citet[][their table\,4]{Baraffe++08aa} find planetary radii of 0.97--1.06\Rjup\ for planets of mass 0.5--1.0\Mjup\ and age 0.5--5\,Gyr, again in accord with our results. These values are for a heavy-element fraction of $Z = 0.02$, and larger fractions result in progressively smaller radii and thus poorer agreement with our radius measurement.

Finally, the properties of WASP-57\,b match the predictions of \citet[][their fig.\,6]{Fortney++07apj} for a 25\,M$_\oplus$ heavy-element core. The difference in radius between models with and without this core are only 0.05\Rjup\ for a 1\Mjup\ planet and 0.18\Rjup\ for a 0.3\Mjup\ planet, so are of a comparable size to the uncertainty in the radius of WASP-57\,b. Our measured properties for this planet therefore do not provide significant support for a high metallicity or the presence of a heavy-element core.


\section{Variation of radius with wavelength}                                                                                       \label{sec:rayleigh}

Two of our datasets include observations in four passbands simultaneously ($ugrz$ for BUSCA and $griz$ for GROND), whereas we have $r$- or $R$-band photometry from four different sources (DFOSC, EulerCam, BUSCA and GROND), so it is relevant to search for possible changes in the measured radius of the planet as a function of wavelength. Such analyses are the photometric equivalent of transmission spectroscopy \citep{SeagerSasselov00apj} and have been pioneered at optical wavelengths by \citet{Sing+11aa}, \citet{Demooij+12aa} and \citet{Me+12mn2}.

Changes in the radius measured from planetary transits, as a function of wavelength, are predicted to occur due to opacity variations which affect the height at which the atmosphere transmits light coming from the parent star in the direction of the observer. At blue wavelengths a greater atmospheric opacity due to Rayleigh and Mie scattering leads to a higher maximum depth at which starlight is transmitted, causing an increase in the measured radius of the planet \citep[e.g.][]{Pont+08mn,Nikolov+15mn}. Enhanced opacity also leads to signatures of sodium and potassium at optical wavelengths \citep{Fortney+08apj}, although only narrow absorption cores have been detected so far \citep{Nikolov+15mn}.

\begin{figure} \includegraphics[width=\columnwidth,angle=0]{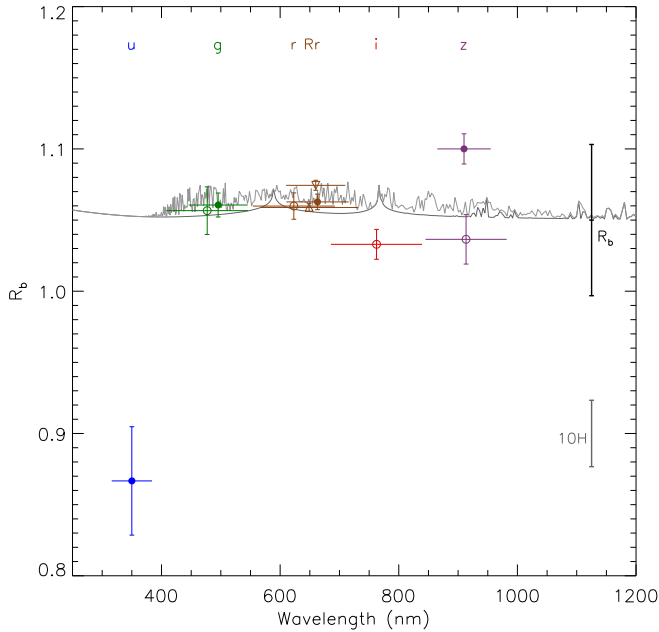}
\caption{\label{fig:rvary} Measured planetary radius ($R_{\rm b}$) as a function of the central
wavelength of the passbands used for the different light curves. The datapoints show the $R_{\rm b}$
measured from each light curve. The vertical errorbars show the relative uncertainty in $R_{\rm b}$
(i.e.\ neglecting the common sources of error) and the horizontal errorbars indicate the FWHM of
the passband. The datapoints are colour-coded according to passband, and the passbands are labelled
at the top of the figure. The symbol types are filled circles (BUSCA), open circles (GROND),
upward-pointing arrow (DFOSC) and downward-pointing arrow (Euler). On the right of the plot we
show the value of $R_{\rm b}$ measured in Section\,\ref{sec:lc} and the size of ten atmospheric
pressure scale heights ($10H$). The grey lines through the empirical datapoints show theoretical
predictions for a transmission spectrum of a gas-giant planet of solar chemical composition from
Madhusudhan (priv.\ comm.). The darker-grey line includes features due to Na and K whereas the
lighter grey line also includes TiO opacity.} \end{figure}

The fundamental observable in this work is the transit depth, represented in our notation by the ratio of the radii $k$ or the fractional planetary radius $r_{\rm b}$. The parameter directly comparable to theoretical predictions is the true planetary radius, $R_{\rm b}$. The parameter $r_{\rm b}$ is correlated with other photometric parameters (see e.g.\ \citealt{Me08mn}), and its transformation into $R_{\rm b}$ requires other parameters which have uncertainty but are common to all photometric passbands (the error budgets calculated in the previous section show that these are $r_{\rm A}$, \Teff\ and \FeH).

We removed these two effects in order to determine $R_{\rm b}$ values with relative errorbars. We did this by refitting the light curves with all parameters fixed except $k$, $T_0$, \reff{the linear LD coefficient for the quadratic LD law,} and the coefficients of the polynomials between differential magnitude and time. We then transformed the resulting $r_{\rm b}$ values into $R_{\rm b}$ using a fixed orbital semimajor axis, $a$. This yielded a set of $R_{\rm b}$ values and errorbars which are directly comparable to each other. The uncertainties in $r_{\rm b}$ were measured using 1000 Monte Carlo simulations each.

\begin{figure} \includegraphics[width=\columnwidth,angle=0]{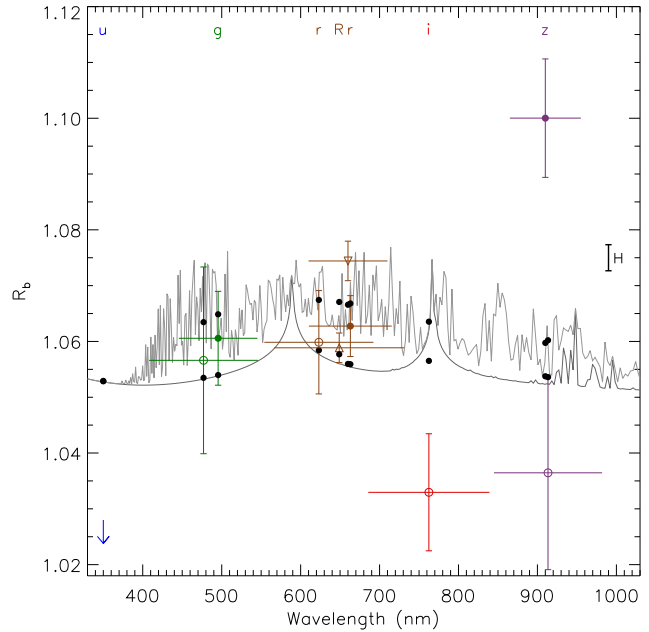}
\caption{\label{fig:rvaryzoom} Close-up of the main part of Fig.\,\ref{fig:rvary} showing
the $R_{\rm b}$ measurements and theoretical transmission spectra. The $u$-band result is
off the plot. Its central wavelength is indicated with a downward-pointing arrow. \reff{The
size of one atmopsheric scale height is indicated to the right of the plot. The black
circles are the values of passband averages of the two transmission spectra, and are shown
at the central wavelengths of the relevant passbands.}} \end{figure}

The data included in this analysis were the two Euler light curves (modelled simultaneously), the two DFOSC light curves (modelled simultaneously), the BUSCA $ugrz$ and the GROND $griz$ data. We did not include the TRAPPIST light curves because the very wide passbands ($I$+$z$ or blue-blocking filters) yield minimal spectral resolution. Although the GROND data only partially cover a transit, which precluded their use in Section\,\ref{sec:lc}, they give reliable results here because $r_{\rm A}$ and $i$ were fixed during the fitting process.

Fig.\,\ref{fig:rvary} shows the resulting values of $R_{\rm b}$ as a function of the central wavelength of the passbands used. The FWHMs of the passbands are shown for reference using horizontal lines. The originating $r_{\rm b}$ values and passband characteristics are collected in Table\,\ref{tab:rb}. Two conclusions are immediately apparent from this figure. Firstly, the planetary radius in the $u$-band is very small and very uncertain. Secondly, the $grRiz$ results are consistent with no variation of $R_{\rm B}$ with wavelength.

The $u$-band light curve shows a small transit depth and a high scatter (see Fig.\,\ref{fig:rvary}), which causes the anomalous $r_{\rm b}$ measurement in this band. The discrepancy relative to the overall $r_{\rm b}$ value obtained in Section\,\ref{sec:lc} is highly significant at 4.8$\sigma$, and corresponds to approximately $39H$. \reff{$H$ is the pressure scale height, and in the case of WASP-57\,b is $334 \pm 35$\,km ($0.0047 \pm 0.0005$\Rjup)}. A variation in $R_{\rm b}$ of the size of $39H$ is difficult to explain, and is not believable unless confirmed by additional data of much higher quality. The two $z$-band light curves also show a clear disagreement of size 3.2$\sigma$.

Fig.\,\ref{fig:rvary} also shows two theoretical transmission spectra calculated under different assumptions by Madhusudhan (priv.\ comm.) using the atmosphere code of \citet{MadhusudhanSeager09apj}. These predictions are for a gas-giant planet of radius 1.25\Rjup\ and surface gravity 25\mss\ so have been scaled to match the smaller radius and lower gravity of WASP-57\,b. Our finding of a small $u$-band $r_{\rm b}$ is not consistent with these theoretical predictions.

\reff{A close-up of the main part of Fig.\,\ref{fig:rvary} is shown in Fig.\,\ref{fig:rvaryzoom}. Passband-averaged values for the transmission spectra are shown with black filled circles, and differ by up to $2H$. The Rayleigh scattering slope could be significantly greater than this: \citet{Sing+11mn} found that Rayleigh scattering caused the measured radius of HD\,189733\,b to be larger by $5H$ at 400\,nm than at 900\,nm.}

\reff{The relative uncertainties in our measured radii for WASP-57\,b are below $1H$ for two, and below $2H$ for five, of the ten light curves (see Table\,\ref{tab:rb}). We are therefore sensitive to radius variations at the level of $1H$, which is smaller than both the difference between the two theoretical transmission spectra and the size of the Rayleigh scattering slope detected for HD\,189733\,b. Our data are therefore sensitive, in principle, to the atmospheric properties of WASP-57\,b. However, in practise, our measurements are insufficient for studying the atmosphere of this planet due to the anomalous result for the $u$-band and the scatter of the radius measurements in the $i$- and $z$-bands. The situation could be improved by obtaining data in narrower passbands (i.e.\ higher spectral resolution), and with repeated observations over the full optical wavelength range. Particular attention should be paid to the $u$-band, which is an important discriminant between the two transmission spectra and also enhances sensitivity to the Rayleigh scattering slope.}


\section{Summary and conclusions}                                                                                                 \label{sec:summary}

WASP-57\,b is a relatively low-mass hot Jupiter orbiting a cool star. Amateur astronomers first noticed that its transits were occurring earlier than predicted, a finding subsequently confirmed by observations from professional facilities. We have presented ten transit light curves from amateur astronomers, plus 13 obtained using professional telescopes of which seven predate the discovery of inaccuracy in the orbital ephemeris of the system. We have determined a revised orbital ephemeris which differs by 24$\sigma$ from the orbital period in the discovery paper, and can be used to predict transits to a precision of less than 1 minute until the year 2170. We also obtained high-resolution Lucky Imaging observations, which show no evidence for nearby companions whose flux might have contaminated our light curves.

\reff{We have used these and previously published data to redetermine the physical properties of the WASP-57 system, finding that both the planet and its host star are larger and less massive than previously thought. A comparison of our new results for WASP-57\,b to theoretical predictions for the properties of gaseous planets reveals a good agreement with models lacking a core or additional heat sources. This disagrees with the core mass of 50\,M$_\oplus$ postulated by F13, but is in accord with expectations for a planet which formed around a star of significantly sub-solar metal abundance.}

We observed two of the transits of WASP-57 using two 2.2\,m telescopes equipped with simultaneous multi-band imaging instruments: GROND ($griz$ passbands) and BUSCA ($ugrz$ passbands). These data are well-suited to investigating the possible wavelength-dependence of the planet's measured radius due to effects such as Rayleigh and Mie scattering, and atomic and molecular absorption. Whilst the radii in the $g$ and $r$/$R$ bands are in generally good agreement, the $i$-band measurement is slightly smaller than expected and the two $z$-band measurements are discrepant by 3$\sigma$. The $u$-band radius is crucial for measuring the Rayleigh scattering slope, as well as separating the pM and pL classes proposed by \citet{Fortney+08apj}. Our measurement is 5$\sigma$ below theoretical predictions, and the size of the discrepancy is inexplicable using current theoretical transmission spectra. This result is almost certainly spurious, and can plausibly be blamed on the strong absorption by Earth's atmosphere at blue-optical wavelengths plus the faintness of the host star in this passband.

\subsection{Future opportunities for transmission photometry}

\reff{Successful detections of radius variations in optical transmission photometry have recently been announced for the TEPs GJ\,3470\,b \citep{Nascimbeni+13aa2,Biddle+14mn}, Qatar-2\,b \citep{Mancini+14mn} and WASP-103\,b \citep{Me+15mn}. Only one of these studies presented data obtained shortward of the Balmer jump, which is an important but observationally difficult wavelength interval (see Fig.\,\ref{fig:rvaryzoom}).}

Transit light curves in the $u$ and $U$ bands have previously been presented for several TEPs, and have shown planetary radii either consistent with other optical passbands (WASP-12, \citealt{Copperwheat+13mn}; TrES-3, \citealt{Turner+13mn}; WASP-17, \citealt{Bento+14mn}; WASP-39 and WASP-43, \citealt{Ricci+15pasp}; XO-2, \citealt{Zellem+15xxx}) or somewhat larger than other optical passbands (HAT-P-5, \citealt{Me+12mn2}, J.\ Dittman 2012, priv.\ comm.; GJ\,3470, \citealt{Nascimbeni+13aa2}). A universal feature of these studies is the reliance on either a single $u$-band transit light curve, which yields large uncertainties on the measured planetary radius, or the use of data not obtained simultaneously in multiple passbands, so the results are hostage to temporal changes such as induced by magnetic activity in the host stars. Most transmission photometry studies also suffer from the use of wide passbands, which are insensitive to spectral features other than broad continuum slopes \citep[see][]{Nikolov+13aa}.

\reff{Whilst suffering from a lower spectral resolution, transmission photometry has several advantages over transmission spectroscopy. These include being able to observe over a wide wavelength interval without being subject to second-order contamination, the ability to use comparison stars more distant from the planet host star, and the option to use telescope defocussing techniques to avoid systematic noise \citep[but see][for a counter-example]{Burton+15mn}. Smaller telescopes can be used, making it easier in particular to observe multiple transits and thus demonstrate the repeatability of the experiment \citep{Bean+13apj,Gibson14mn}.}

We therefore advocate studies based on observations of multiple transits, obtained simultaneously through many intermediate or narrow passbands. These passbands should be well-defined by interference filters, thus avoiding compromises such as the variable red edge of the $z$ filter due to its reliance on the quantum efficiency curve of the CCD used \citep[e.g.][]{Fukugita+96aj} or the red leak in some $u$ and $U$ filters which is capable of causing spurious results for optically-blue objects \citep[e.g.][]{Guhathakurta+98aj}. With a sufficient number of passbands, it should be possible to achieve low-resolution spectroscopy of the atmospheres of extrasolar planets through the full optical wavelength range using the transmission-photometry approach.


\section*{Acknowledgements}

Giorgio Corfini suddenly passed away at the end of 2014. He had been for many years an active observer and member of the ``Unione Astrofili Italiani'' (UAI). The UAI and the working groups of the sections ``Extrasolar Planets'' and ``Variable Stars'' acknowledge his important contribution and would like to dedicate this paper to his memory.

The operation of the Danish 1.54m telescope is financed by a grant to UGJ from the Danish Natural Science Research Council (FNU).
This paper incorporates observations collected using the Gamma Ray Burst Optical and Near-Infrared Detector (GROND) instrument at the MPG 2.2\,m telescope located at ESO La Silla, Chile, program 093.A-9007(A). GROND was built by the high-energy group of MPE in collaboration with the LSW Tautenburg and ESO, and is operated as a PI-instrument at the MPG 2.2\,m telescope. 
This paper incorporates observations collected at the Centro Astron\'omico Hispano Alem\'an (CAHA) at Calar Alto, Spain, operated jointly by the Max-Planck Institut f\"ur Astronomie and the Instituto de Astrof\'{\i}sica de Andaluc\'{\i}a (CSIC).
TRAPPIST is funded by the Belgian Fund for Scientific Research (Fond National de la Recherche Scientifique, FNRS) under the grant FRFC 2.5.594.09.F, with the participation of the Swiss National Science Fundation (SNF). MG and EJ are FNRS Research Associates. LD is a FNRS/FRIA Doctoral Fellow.
We thank the anonymous referee for a helpful report and Dr.\ Francesca Faedi for discussions.
The reduced light curves presented in this work will be made available at the CDS ({\tt http://vizier.u-strasbg.fr/}) and at {\tt http://www.astro.keele.ac.uk/$\sim$jkt/}.
J\,Southworth acknowledges financial support from STFC in the form of an Advanced Fellowship.
%
%
%
%
This publication was partially supported by grant NPRP X-019-1-006 from Qatar National Research Fund (a member of Qatar Foundation).
TCH is supported by the Korea Astronomy \& Space Science Institute travel grant \#2014-1-400-06.
%
%
%
%
TCH acknowledges support from the Korea Astronomy and Space Science Institute (KASI) grant 2014-1-400-06.
%
%
%
OW (FNRS research fellow) and J\,Surdej acknowledge support from the Communaut\'e fran\c{c}aise de Belgique - Actions de recherche concert\'ees - Acad\'emie Wallonie-Europe.
The following internet-based resources were used in research for this paper: the ESO Digitized Sky Survey; the NASA Astrophysics Data System; the SIMBAD database and VizieR catalogue access tool operated at CDS, Strasbourg, France; and the ar$\chi$iv scientific paper preprint service operated by Cornell University.

\bibliographystyle{mn_new}

\begin{thebibliography}{69}
\expandafter\ifx\csname natexlab\endcsname\relax\def\natexlab#1{#1}\fi

\bibitem[{{Baraffe} et~al.(2008){Baraffe}, {Chabrier}, \&
  {Barman}}]{Baraffe++08aa}
{Baraffe}, I., {Chabrier}, G., {Barman}, T., 2008, A\&A, 482, 315

\bibitem[{{Bean} et~al.(2013){Bean}, {D{\'e}sert}, {Seifahrt}, {Madhusudhan},
  {Chilingarian}, {Homeier}, \& {Szentgyorgyi}}]{Bean+13apj}
{Bean}, J.~L., {D{\'e}sert}, J.-M., {Seifahrt}, A., {Madhusudhan}, N.,
  {Chilingarian}, I., {Homeier}, D., {Szentgyorgyi}, A., 2013, ApJ, 771, 108

\bibitem[{{Ben{\'{\i}}tez-Llambay} et~al.(2011){Ben{\'{\i}}tez-Llambay},
  {Masset}, \& {Beaug{\'e}}}]{Benitez++11aa}
{Ben{\'{\i}}tez-Llambay}, P., {Masset}, F., {Beaug{\'e}}, C., 2011, A\&A, 528,
  A2

\bibitem[{{Bento} et~al.(2014)}]{Bento+14mn}
{Bento}, J., et~al., 2014, MNRAS, 437, 1511

\bibitem[{{Biddle} et~al.(2014)}]{Biddle+14mn}
{Biddle}, L.~I., et~al., 2014, MNRAS, 443, 1810

\bibitem[{{Bodenheimer} et~al.(2003){Bodenheimer}, {Laughlin}, \&
  {Lin}}]{Bodenheimer++03apj}
{Bodenheimer}, P., {Laughlin}, G., {Lin}, D.~N.~C., 2003, ApJ, 592, 555

\bibitem[{{Buchhave} et~al.(2014)}]{Buchhave+14nat}
{Buchhave}, L.~A., et~al., 2014, Nature, 509, 593

\bibitem[{{Burton} et~al.(2015){Burton}, {Watson}, {Rodr{\'{\i}}guez-Gil},
  {Skillen}, {Littlefair}, {Dhillon}, \& {Pollacco}}]{Burton+15mn}
{Burton}, J.~R., {Watson}, C.~A., {Rodr{\'{\i}}guez-Gil}, P., {Skillen}, I.,
  {Littlefair}, S.~P., {Dhillon}, S., {Pollacco}, D., 2015, MNRAS, 446, 1071

\bibitem[{{Carter} et~al.(2008){Carter}, {Yee}, {Eastman}, {Gaudi}, \&
  {Winn}}]{Carter+08apj}
{Carter}, J.~A., {Yee}, J.~C., {Eastman}, J., {Gaudi}, B.~S., {Winn}, J.~N.,
  2008, ApJ, 689, 499

\bibitem[{{Charbonneau} et~al.(2000){Charbonneau}, {Brown}, {Latham}, \&
  {Mayor}}]{Charbonneau+00apj}
{Charbonneau}, D., {Brown}, T.~M., {Latham}, D.~W., {Mayor}, M., 2000, ApJ,
  529, L45

\bibitem[{{Claret}(2004)}]{Claret04aa}
{Claret}, A., 2004, A\&A, 424, 919

\bibitem[{{Copperwheat} et~al.(2013)}]{Copperwheat+13mn}
{Copperwheat}, C.~M., et~al., 2013, MNRAS, 434, 661

\bibitem[{{Daemgen} et~al.(2009){Daemgen}, {Hormuth}, {Brandner}, {Bergfors},
  {Janson}, {Hippler}, \& {Henning}}]{Daemgen+09aa}
{Daemgen}, S., {Hormuth}, F., {Brandner}, W., {Bergfors}, C., {Janson}, M.,
  {Hippler}, S., {Henning}, T., 2009, A\&A, 498, 567

\bibitem[{{de Mooij} et~al.(2012)}]{Demooij+12aa}
{de Mooij}, E.~J.~W., et~al., 2012, A\&A, 538, A46

\bibitem[{{Demarque} et~al.(2004){Demarque}, {Woo}, {Kim}, \&
  {Yi}}]{Demarque+04apjs}
{Demarque}, P., {Woo}, J.-H., {Kim}, Y.-C., {Yi}, S.~K., 2004, ApJS, 155, 667

\bibitem[{{Dominik} et~al.(2010)}]{Dominik+10an}
{Dominik}, M., et~al., 2010, AN, 331, 671

\bibitem[{{Dotter} et~al.(2008){Dotter}, {Chaboyer}, {Jevremovi{\'c}},
  {Kostov}, {Baron}, \& {Ferguson}}]{Dotter+08apjs}
{Dotter}, A., {Chaboyer}, B., {Jevremovi{\'c}}, D., {Kostov}, V., {Baron}, E.,
  {Ferguson}, J.~W., 2008, ApJS, 178, 89

\bibitem[{{Eastman} et~al.(2010){Eastman}, {Siverd}, \&
  {Gaudi}}]{Eastman++10pasp}
{Eastman}, J., {Siverd}, R., {Gaudi}, B.~S., 2010, PASP, 122, 935

\bibitem[{{Faedi} et~al.(2013)}]{Faedi+13aa}
{Faedi}, F., et~al., 2013, A\&A, 551, A73

\bibitem[{{Fortney} et~al.(2007){Fortney}, {Marley}, \&
  {Barnes}}]{Fortney++07apj}
{Fortney}, J.~J., {Marley}, M.~S., {Barnes}, J.~W., 2007, ApJ, 659, 1661

\bibitem[{{Fortney} et~al.(2008){Fortney}, {Lodders}, {Marley}, \&
  {Freedman}}]{Fortney+08apj}
{Fortney}, J.~J., {Lodders}, K., {Marley}, M.~S., {Freedman}, R.~S., 2008, ApJ,
  678, 1419

\bibitem[{{Fukugita} et~al.(1996){Fukugita}, {Ichikawa}, {Gunn}, {Doi},
  {Shimasaku}, \& {Schneider}}]{Fukugita+96aj}
{Fukugita}, M., {Ichikawa}, T., {Gunn}, J.~E., {Doi}, M., {Shimasaku}, K.,
  {Schneider}, D.~P., 1996, AJ, 111, 1748

\bibitem[{{Gibson}(2014)}]{Gibson14mn}
{Gibson}, N.~P., 2014, MNRAS, 445, 3401

\bibitem[{{Gibson} et~al.(2009)}]{Gibson+09apj}
{Gibson}, N.~P., et~al., 2009, ApJ, 700, 1078

\bibitem[{{Gillon} et~al.(2011){Gillon}, {Jehin}, {Magain}, {Chantry},
  {Hutsem{\'e}kers}, {Manfroid}, {Queloz}, \& {Udry}}]{Gillon+11conf}
{Gillon}, M., {Jehin}, E., {Magain}, P., {Chantry}, V., {Hutsem{\'e}kers}, D.,
  {Manfroid}, J., {Queloz}, D., {Udry}, S., 2011, in European Physical Journal
  Web of Conferences, vol.~11, p. 6002

\bibitem[{{Gillon} et~al.(2013)}]{Gillon+13aa}
{Gillon}, M., et~al., 2013, A\&A, 552, A82

\bibitem[{{Greiner} et~al.(2008)}]{Greiner+08pasp}
{Greiner}, J., et~al., 2008, PASP, 120, 405

\bibitem[{{Guhathakurta} et~al.(1998){Guhathakurta}, {Webster}, {Yanny},
  {Schneider}, \& {Bahcall}}]{Guhathakurta+98aj}
{Guhathakurta}, P., {Webster}, Z.~T., {Yanny}, B., {Schneider}, D.~P.,
  {Bahcall}, J.~N., 1998, AJ, 116, 1757

\bibitem[{{Hellier} et~al.(2012)}]{Hellier+12mn}
{Hellier}, C., et~al., 2012, MNRAS, 426, 739

\bibitem[{{Henry} et~al.(2000){Henry}, {Marcy}, {Butler}, \&
  {Vogt}}]{Henry+00apj}
{Henry}, G.~W., {Marcy}, G.~W., {Butler}, R.~P., {Vogt}, S.~S., 2000, ApJ, 529,
  L41

\bibitem[{{Jehin} et~al.(2011)}]{Jehin+11msngr}
{Jehin}, E., et~al., 2011, The Messenger, 145, 2

\bibitem[{{Konacki} et~al.(2003){Konacki}, {Torres}, {Jha}, \&
  {Sasselov}}]{Konacki+03nat}
{Konacki}, M., {Torres}, G., {Jha}, S., {Sasselov}, D.~D., 2003, Nature, 421,
  507

\bibitem[{{Lendl} et~al.(2012)}]{Lendl+12aa}
{Lendl}, M., et~al., 2012, A\&A, 544, A72

\bibitem[{{Lucy} \& {Sweeney}(1971)}]{LucySweeney71aj}
{Lucy}, L.~B., {Sweeney}, M.~A., 1971, AJ, 76, 544

\bibitem[{{Madhusudhan} \& {Seager}(2009)}]{MadhusudhanSeager09apj}
{Madhusudhan}, N., {Seager}, S., 2009, ApJ, 707, 24

\bibitem[{{Mancini} et~al.(2014)}]{Mancini+14mn}
{Mancini}, L., et~al., 2014, MNRAS, 443, 2391

\bibitem[{{Mordasini} et~al.(2009{\natexlab{a}}){Mordasini}, {Alibert}, \&
  {Benz}}]{Mordasini+09aa}
{Mordasini}, C., {Alibert}, Y., {Benz}, W., 2009{\natexlab{a}}, A\&A, 501, 1139

\bibitem[{{Mordasini} et~al.(2009{\natexlab{b}}){Mordasini}, {Alibert}, {Benz},
  \& {Naef}}]{Mordasini+09aa2}
{Mordasini}, C., {Alibert}, Y., {Benz}, W., {Naef}, D., 2009{\natexlab{b}},
  A\&A, 501, 1161

\bibitem[{{Nascimbeni} et~al.(2013){Nascimbeni}, {Piotto}, {Pagano},
  {Scandariato}, {Sani}, \& {Fumana}}]{Nascimbeni+13aa2}
{Nascimbeni}, V., {Piotto}, G., {Pagano}, I., {Scandariato}, G., {Sani}, E.,
  {Fumana}, M., 2013, A\&A, 559, A32

\bibitem[{{Nikolov} et~al.(2013){Nikolov}, {Chen}, {Fortney}, {Mancini},
  {Southworth}, {van Boekel}, \& {Henning}}]{Nikolov+13aa}
{Nikolov}, N., {Chen}, G., {Fortney}, J., {Mancini}, L., {Southworth}, J., {van
  Boekel}, R., {Henning}, T., 2013, A\&A, 553, A26

\bibitem[{{Nikolov} et~al.(2015)}]{Nikolov+15mn}
{Nikolov}, N., et~al., 2015, MNRAS, 447, 463

\bibitem[{{Pietrinferni} et~al.(2004){Pietrinferni}, {Cassisi}, {Salaris}, \&
  {Castelli}}]{Pietrinferni+04apj}
{Pietrinferni}, A., {Cassisi}, S., {Salaris}, M., {Castelli}, F., 2004, ApJ,
  612, 168

\bibitem[{{Poddan{\'y}} et~al.(2010){Poddan{\'y}}, {Br{\'a}t}, \&
  {Pejcha}}]{Poddany++10newa}
{Poddan{\'y}}, S., {Br{\'a}t}, L., {Pejcha}, O., 2010, New Astronomy, 15, 297

\bibitem[{{Pollacco} et~al.(2006)}]{Pollacco+06pasp}
{Pollacco}, D.~L., et~al., 2006, PASP, 118, 1407

\bibitem[{{Pont} et~al.(2008){Pont}, {Knutson}, {Gilliland}, {Moutou}, \&
  {Charbonneau}}]{Pont+08mn}
{Pont}, F., {Knutson}, H., {Gilliland}, R.~L., {Moutou}, C., {Charbonneau}, D.,
  2008, MNRAS, 385, 109

\bibitem[{{Ricci} et~al.(2015)}]{Ricci+15pasp}
{Ricci}, D., et~al., 2015, PASP, 127, 143

\bibitem[{{Rowe} et~al.(2014)}]{Rowe+14apj}
{Rowe}, J.~F., et~al., 2014, ApJ, 784, 45

\bibitem[{{Seager} \& {Mall{\'e}n-Ornelas}(2003)}]{SeagerMallen03apj}
{Seager}, S., {Mall{\'e}n-Ornelas}, G., 2003, ApJ, 585, 1038

\bibitem[{{Seager} \& {Sasselov}(2000)}]{SeagerSasselov00apj}
{Seager}, S., {Sasselov}, D.~D., 2000, ApJ, 537, 916

\bibitem[{{Sing} et~al.(2011{\natexlab{a}})}]{Sing+11aa}
{Sing}, D.~K., et~al., 2011{\natexlab{a}}, A\&A, 527, A73

\bibitem[{{Sing} et~al.(2011{\natexlab{b}})}]{Sing+11mn}
{Sing}, D.~K., et~al., 2011{\natexlab{b}}, MNRAS, 416, 1443

\bibitem[{{Skottfelt} et~al.(2013)}]{Skottfelt+13aa}
{Skottfelt}, J., et~al., 2013, A\&A, 553, A111

\bibitem[{{Skottfelt} et~al.(2015)}]{Skottfelt+15aa}
{Skottfelt}, J., et~al., 2015, A\&A, 574, A54

\bibitem[{{Southworth}(2008)}]{Me08mn}
{Southworth}, J., 2008, MNRAS, 386, 1644

\bibitem[{{Southworth}(2009)}]{Me09mn}
{Southworth}, J., 2009, MNRAS, 394, 272

\bibitem[{{Southworth}(2010)}]{Me10mn}
{Southworth}, J., 2010, MNRAS, 408, 1689

\bibitem[{{Southworth}(2011)}]{Me11mn}
{Southworth}, J., 2011, MNRAS, 417, 2166

\bibitem[{{Southworth}(2012)}]{Me12mn}
{Southworth}, J., 2012, MNRAS, 426, 1291

\bibitem[{{Southworth} et~al.(2004){Southworth}, {Maxted}, \&
  {Smalley}}]{Me++04mn}
{Southworth}, J., {Maxted}, P.~F.~L., {Smalley}, B., 2004, MNRAS, 349, 547

\bibitem[{{Southworth} et~al.(2007){Southworth}, {Wheatley}, \&
  {Sams}}]{Me++07mn}
{Southworth}, J., {Wheatley}, P.~J., {Sams}, G., 2007, MNRAS, 379, L11

\bibitem[{{Southworth} et~al.(2012){Southworth}, {Mancini}, {Maxted}, {Bruni},
  {Tregloan-Reed}, {Barbieri}, {Ruocco}, \& {Wheatley}}]{Me+12mn2}
{Southworth}, J., {Mancini}, L., {Maxted}, P.~F.~L., {Bruni}, I.,
  {Tregloan-Reed}, J., {Barbieri}, M., {Ruocco}, N., {Wheatley}, P.~J., 2012,
  MNRAS, 422, 3099

\bibitem[{{Southworth} et~al.(2009)}]{Me+09mn}
{Southworth}, J., et~al., 2009, MNRAS, 396, 1023

\bibitem[{{Southworth} et~al.(2014)}]{Me+14mn}
{Southworth}, J., et~al., 2014, MNRAS, 444, 776

\bibitem[{{Southworth} et~al.(2015)}]{Me+15mn}
{Southworth}, J., et~al., 2015, MNRAS, 447, 711

\bibitem[{{Stetson}(1987)}]{Stetson87pasp}
{Stetson}, P.~B., 1987, PASP, 99, 191

\bibitem[{{Turner} et~al.(2013)}]{Turner+13mn}
{Turner}, J.~D., et~al., 2013, MNRAS, 428, 678

\bibitem[{{VandenBerg} et~al.(2006){VandenBerg}, {Bergbusch}, \&
  {Dowler}}]{Vandenberg++06apjs}
{VandenBerg}, D.~A., {Bergbusch}, P.~A., {Dowler}, P.~D., 2006, ApJS, 162, 375

\bibitem[{{Wang} \& {Fischer}(2015)}]{WangFischer15aj}
{Wang}, J., {Fischer}, D.~A., 2015, AJ, 149, 14

\bibitem[{{Zellem} et~al.(2015)}]{Zellem+15xxx}
{Zellem}, R.~T., et~al., 2015, ApJ, 810, 11

\end{thebibliography}

\end{document}